\documentclass[10pt]{IEEEtran}
\usepackage{amsthm,amsmath,comment,bbm,amssymb,color,graphicx}
\usepackage{mathtools}
\usepackage{comment}
\usepackage{url}
\usepackage{caption}
\usepackage{subcaption}
\usepackage{epstopdf}
\usepackage{longtable}
\usepackage{cite}
\usepackage{tikz}
\usepackage{multirow}
\usepackage{multicol}

\def\nbh{{\mathbf{h}}}

\def\nbp{{\mathbf{p}}}

\def\nbx{{\mathbf{x}}}

\def\nb0{{\mathbf{0}}}
\def\nb1{{\mathbf{1}}}

\def\nbH{{\mathbf{H}}}

\def\nbP{{\mathbf{P}}}

\def\ncalM{{\mathcal{M}}}

\def\nbbC{{\mathbb{C}}}

\def\nbbM{{\mathbb{M}}}

\def\nbbP{{\mathbb{P}}}

\newtheorem{nrem}{Remark}

\newtheorem{eg}{Example}

\begin{document}
\title{Rate-Splitting Multiple Access: The First Prototype and Experimental Validation of its Superiority over SDMA and NOMA}
\author{Xinze Lyu,~\IEEEmembership{Student Member, IEEE}, Sundar Aditya,~\IEEEmembership{Member, IEEE}, Junghoon Kim,~\IEEEmembership{Member, IEEE} and Bruno Clerckx,~\IEEEmembership{Fellow, IEEE}
\thanks{This work was supported in part by UKRI Impact Acceleration Account (IAA) grant EP/X52556X/1.}
\thanks{X.~Lyu, S.~Aditya and B.~Clerckx are with the Dept.~of Electrical and Electronic Engg., Imperial College London, London SW7 2AZ, U.K. (e-mail: \{x.lyu21, s.aditya, b.clerckx\}@imperial.ac.uk). B.~Clerckx is also with Silicon Austria Labs, 8010 Graz, Austria.}
\thanks{J.~Kim is with the College of Ocean Science and Engineering, National Korea Maritime \& Ocean University, Busan 49112, South Korea (e-mail: j.kim@kmou.ac.kr).}
}
\maketitle

\begin{abstract}
In multi-user multi-antenna communications, it is well-known in theory that Rate-Splitting Multiple Access (RSMA) can achieve a higher spectral efficiency than both Space Division Multiple Access (SDMA) and Non-Orthogonal Multiple Access (NOMA). However, an experimental evaluation of RSMA's performance, relative to SDMA and NOMA, is missing in the literature, which is essential to address the ongoing debate between RSMA and NOMA over which is better suited to handle most efficiently the available resources and interference in 6G. In this paper, we address this critical knowledge gap by realizing the first-ever RSMA prototype using software-defined radios. Through measurements using our prototype, we empirically solve the modulation and coding scheme limited sum throughput maximization problem for RSMA, SDMA and NOMA for the two-user multiple-input single-output (MISO) scenario over (a) different pairs of line-of-sight channels that vary in terms of their relative pathloss and spatial correlation, and with (b) different channel state information quality. We observe that RSMA achieves the highest sum throughput across all these cases, whereas SDMA and NOMA are effective only in some cases. Furthermore, RSMA also achieves better fairness at a higher sum throughput than both SDMA and NOMA.

\end{abstract}


\bstctlcite{IEEEexample:BSTcontrol}

\section{Introduction}
The evolutionary trajectory over successive generations of wireless networks has been towards achieving higher spectral efficiency, greater user fairness and higher energy efficiency. For upcoming 6G in particular, improvements in spectral and energy efficiency of at least 2x over 5G have been speculated \cite{Samsung6G}. To meet these targets, the underlying multiple access technique must be capable of offering higher efficiency, flexibility and coping with much higher levels of multi-user interference than Space Division Multiple Access (SDMA) -- the state-of-the-art multiple access technique used in current wireless standards such as 3GPP 5G NR and IEEE 802.11. The two candidates widely believed to be capable of achieving this in future wireless networks are Non-Orthogonal Multiple Access (NOMA) and Rate-Splitting Multiple Access (RSMA). We provide a brief description of each below, specifically focusing on practical implementations.

\subsection{NOMA}
NOMA is inspired by the fact that the capacity region of the $K$-user single-input-single-output (SISO) broadcast channel is achieved by superposition coding at the transmitter (TX) and successive interference cancellation (SIC) at the receivers (RXs) \cite{siso_bc}. In SISO NOMA, users are ordered in decreasing order of their channel strength and the $l$-th strongest user decodes and subtracts the interference from the $K-l$ weaker users (i.e., SIC) before decoding its desired signal. The interference from the $l-1$ stronger users is treated as noise\footnote{This is commonly referred to as power-domain NOMA to distinguish it from other variants, such as code-domain and frequency-domain NOMA \cite{Islam_NOMA_survey_2017}. In this paper, NOMA refers only to power-domain NOMA.}.

The optimality of NOMA for the SISO broadcast channel (in terms of achieving capacity) has been the driving force behind extensive investigations into NOMA for multiple-input-multiple-output (MIMO) systems with the aim of improving spectral efficiency \cite{noma_spectrum_efficiency, noma_vs_zflp} and fairness \cite{noma_fairness_pa} in 5G networks and beyond \cite{noma_3gpp_proposal}. However, for the multi-antenna broadcast channel, NOMA is not optimal, in general, in a capacity-achieving sense \cite{NOMAineffcient}. This is because with multiple antennas at the TX, the spatial domain can be used to suppress interference without SIC at the RXs through either multi-user linear precoding (MU-LP) or non-linear precoding at the TX. MU-LP is the common approach adopted by SDMA \cite{mulp_sdma_defacto}. Essentially, SDMA and NOMA represent two distinct interference management strategies; in SDMA, the interference is treated as noise, whereas in NOMA the interference from the weaker users is decoded by the stronger users \cite{Mao2018}. Hence, NOMA provides substantial spectral efficiency gains over SDMA only when MU-LP cannot suppress the interference to near-noise levels. This occurs either in overloaded systems (i.e., more users than the number of antennas at the TX) or when the user channels in underloaded systems experience high spatial correlation. On the other hand, in non-overloaded systems where user channels do not experience high spatial correlation, NOMA can perform worse than SDMA \cite{NOMAineffcient}. While it can be argued that under high user density, groups of users with highly spatially correlated channels can be found with high probability, it does mean that MIMO-NOMA requires a user grouping strategy based on their channels, which has given rise to two approaches - beamforming-based and cluster-based NOMA \cite{mimo_noma_user_grouping}.

\subsection*{NOMA Implementation}
In terms of experimental evaluations, several papers have focussed on SISO-NOMA \cite[Table~1]{Qi_NOMA_experiments_2021}. MIMO-NOMA implementations are relatively fewer \cite{noma_emulator_measurements, noma_outdoor_measurements, noma_mimo_measurements}. Link and system-level simulations were carried out in \cite{noma_emulator_measurements}, along with measurements over emulated fading channels. Measurements involving over-the-air transmissions were carried out in \cite{noma_outdoor_measurements, noma_mimo_measurements}. However, in these papers, the NOMA performance is compared with orthogonal multiple access instead of SDMA. To the best of our knowledge, no experimental comparison between NOMA and SDMA exists in the literature.

\subsection{RSMA}
{ Unlike NOMA with its SISO roots, RSMA was designed for multi-antenna communications from the outset. For the $K$-user multiple-input-single-output (MISO) case, RSMA produces $K+1$ data streams (i.e., channel-coded and modulated symbol streams) at the TX from the $K$ user messages \cite{HamdiMISOImperfectCSIT} (details in Section~\ref{subsec:stage 2}). In contrast, both SDMA and NOMA generate $K$ data streams. Similar to SDMA, each of the $K+1$ streams is linearly precoded (i.e., MU-LP) and transmitted over the air. At each RX, SIC is used (like NOMA) to decode \emph{only two} streams to retrieve the desired message, regardless of a users's channel strength relative to the other users. Hence, unlike NOMA, no user grouping mechanism prior to data transmission is needed to realize spectral efficiency gains over SDMA. 

RSMA's spectral efficiency gains over SDMA and NOMA stem from the fact that it is degrees-of-freedom (DoF) optimal under perfect and imperfect channel state information (CSI) \cite{NOMAineffcient, HamdiMISOImperfectCSIT}. Furthermore, RSMA has also been shown, in theory, to achieve spectral efficiency gains over SDMA and NOMA under varying interference levels, network loads and user deployments \cite{Mao2018}. Apart from spectral efficiency, RSMA has also been shown to achieve -- again, in theory -- performance improvements over SDMA and NOMA in terms of other metrics such as energy efficiency, fairness, mixed-critical quality-of-service and tight latency requirements, etc. (see \cite{mao2022fundmental, RSMA_JSAC_Primer} and the references therein for a comprehensive survey).} These performance improvements stem from RSMA's unique interference mitigation strategy, wherein at each user, the interference is \emph{partially decoded and partially treated as noise}. In doing so, RSMA generalizes both SDMA and NOMA \cite{Mao2018, RSMAUnifying}. {Taken together, all of these features make RSMA an attractive candidate for physical layer multiple access in 6G, provided the promised gains can be demonstrated in practice. In particular, an experimental comparison between RSMA, SDMA and NOMA is needed as a starting point to address the pressing question of which multiple access technique is best equipped to deliver the spectral efficiency enhancements expected from 6G.}

\subsection*{RSMA Implementation}
Despite its impressive benefits, the practical implementation of RSMA is still in its infancy, with the state-of-the-art focusing on the closely related issues of erroneous SIC, receiver design, finite constellation and adaptive modulation and coding (AMC) for RSMA to minimize the block error rate due to the combined effects of imperfect/outdated CSI at the TX (CSIT) and finite blocklength \cite{LinkLvSimulation, LinkLvSimulation_other, zhang2023ratesplitting, mosquera2023linkadaption, flexibleRSMA}. Among these, RSMA link-level simulations were conducted in \cite{LinkLvSimulation, LinkLvSimulation_other, zhang2023ratesplitting}, feedback-based RSMA link adaptation was investigated in \cite{mosquera2023linkadaption}, while \cite{flexibleRSMA} considered a flexible variant of RSMA, where only a subset of the receivers employ SIC. Though important, these practical aspects have also been tackled only through theory/simulations, and no experimental verification of RSMA's benefits has been reported till date. 

{In this paper, we address this critical knowledge gap by realizing the first-ever RSMA prototype using software-defined radios (SDRs). Using our prototype, we experimentally compare the throughput and fairness performance of RSMA, SDMA and NOMA.} Our contributions are as follows:
{
\begin{itemize}
    \item We implement RSMA on SDRs using a two-stage protocol. In Stage 1, orthogonal pilot signals are transmitted by the TX for CSI estimation at each RX. The estimated CSI is then fed back to the TX, either unquantized or quantized to emulate perfect and imperfect CSIT, respectively (Section~\ref{subsec:stage1}). The CSIT is used to design the precoded RSMA signal that is transmitted in Stage 2 (Section~\ref{subsec:stage 2}). We use OFDM signals largely based on IEEE 802.11g physical layer frames to implement the data payload in the RSMA signal (Section~\ref{sec: RSMA SDR prototype}). The above two-stage protocol can also be used to implement SDMA and NOMA in our prototype with appropriate changes to Stage 2.
    
    \item For the two-user MISO scenario, we formulate the practically relevant \emph{modulation and coding scheme (MCS)-limited sum throughput maximization problem} for RSMA, and establish its relationship to the canonical sum rate maximization problem that is widely analyzed in theory (Section~\ref{subsec:stage 2}). The MCS-limited sum throughput maximization problem can also be formulated for SDMA and NOMA (Table~\ref{tab:rsma_noma_sdma}).

    \item Through measurements using our prototype, we empirically solve the MCS-limited sum throughput maximization problem for RSMA, SDMA and NOMA for the two-user MISO scenario over line-of-sight channels in a lab environment, where the pair of channels differ in terms of their relative strength and spatial correlation (Section~\ref{sec: measurement results}). We empirically observe that 
    
    \begin{itemize}
        \item With unquantized CSI feedback, NOMA achieves significant throughput gains over SDMA when channels have high spatial correlation. On the other hand, when the channels have low spatial correlation, the SDMA throughput is higher than NOMA. RSMA, however, achieves the highest throughput not only under either extreme but also over channels that lie in between w.r.t their spatial correlation. These outcomes are consistent with theoretical predictions in \cite{Mao2018, RSMAUnifying}.
        
        \item With quantized CSI feedback, NOMA achieves significant throughput gains over SDMA even with channels having low spatial correlation, provided there is a sufficiently large disparity between the channel strengths. The NOMA throughput performance is also at par with that of RSMA under these conditions. Essentially, the imperfect CSI disproportionately diminishes the weaker user's throughput for SDMA, whereas both RSMA and NOMA provide higher throughput to the weaker user via the common stream (see Section~\ref{subsec:stage 2} for definition). However, when the channel strength disparity is not sufficiently large, RSMA achieves throughput gains over NOMA as well. These outcomes are consistent with theoretical predictions in \cite{NOMAineffcient, mao2022fundmental, RSMA_JSAC_Primer}.

        \item With both quantized and unquantized CSI feedback, RSMA achieves better fairness at higher throughputs than both SDMA and NOMA. This too is consistent with theoretical predictions in \cite{NOMAineffcient, mao2022fundmental, RSMA_JSAC_Primer}.


    \end{itemize}

\end{itemize}
}

\subsection{Organization}
The rest of this paper is organized as follows. In Section \ref{sec: system model}, {we introduce the RSMA system model for the two-user MISO scenario and formulate the practically relevant MCS-limited sum throughput maximization problem. In Section~\ref{sec: RSMA SDR prototype}, we provide a detailed description of our SDR-based RSMA implementation. In Section~\ref{sec: measurement results}, we empirically solve the MCS-limited sum throughput maximization problem for RSMA, SDMA and NOMA through measurements using our prototype and compare the sum throughput and fairness performance of the three schemes. Finally, Section~\ref{sec:summary} concludes the paper}.

\subsection{Notation}
Column vectors are represented using lowercase bold letters (e.g., $\nbx$). $(\cdot)^H, \|\cdot\|, \cup$ and $\emptyset$ denote the Hermitian operator, the Euclidean norm, the union operation and the empty set, respectively. ${\rm Re}(\cdot)$ and ${\rm Im}(\cdot)$ denote the real and imaginary parts of a complex number, $\lfloor \cdot \rfloor$ denotes the floor function, and $\mathcal{CN}(0,\sigma^2)$ denotes the circularly symmetric complex Gaussian distribution with zero mean and variance $\sigma^2$. 

\section{System Model}
\label{sec: system model}
{
Consider a two-antenna TX communicating with two single-antenna users/RXs, using OFDM\footnote{In principle, RSMA can be used with any waveform. Our choice of OFDM is motivated by its prevalence in modern communications systems. However, in the context of 6G, alternatives to OFDM, such as OTFS have received considerable attention. RSMA for such waveforms -- especially experimental evaluations -- is a relevant topic yet to be investigated.} signals over $N_c$ subcarriers. The communications takes place over two stages.
    \begin{itemize}
        \item \textbf{Stage 1}: First, the TX transmits orthogonal pilot signals for channel estimation at the RXs. The estimated CSI is then fed back to the TX through a separate link.
        
        \item \textbf{Stage 2}: Using the CSI acquired in Stage 1, the TX designs the precoder and the transmit signal to maximize the sum throughput.
    \end{itemize}
We describe each stage, in turn, in Sections \ref{subsec:stage1} and \ref{subsec:stage 2}.
}

\subsection{Stage 1: Channel Model and CSI feedback}
\label{subsec:stage1}
{
Let $\mathbf{h}_i[k] := [h_{i1}[k], ~ h_{i2}[k]]^T \in \nbbC^{2\times 1}$ denote the true frequency flat, slowly varying channel between the TX and RX~$i~(=1,2)$ over the $k$-th subcarrier $(k = 0, \cdots, N_c - 1)$. Similarly, let $\hat{\nbh}_i[k]$ denote the estimate of $\mathbf{h}_i[k]$, obtained at RX~$i$ through pilot signals transmitted by the TX. 
To reduce feedback overhead, let $\hat{\nbh}_i$ denote the \emph{wideband CSI} evaluated at RX~$i$ by averaging $\hat{\nbh}_i[k]$ over the subcarriers, i.e.,    
    \begin{align}
    \label{eq:wideband_csir}
        \hat{\nbh}_i := \frac{1}{N_c} \sum_{k=0}^{N_c-1} \hat{\nbh}_i[k], ~ (i=1, 2).
    \end{align}
For CSI feedback of $\hat{\nbh}_i$ from RX~$i$ to the TX in our experiments, we consider two cases to emulate perfect and imperfect CSI at the TX. 
\begin{itemize}
    \item \emph{Unquantized feedback}: Let $\hat{\nbh}_i^{\rm UQ} = [\hat{h}_{i1}^{\rm UQ} ~ \hat{h}_{i2}^{\rm UQ}]$ denote the unquantized version of $\hat{\nbh}_i$, where each complex $\hat{h}_{il}^{\rm UQ} ~(i,l \in \{1,2\})$ is a 128-bit representation of $\hat{h}_{il}$  (i.e., a 64-bit double-precision floating-point number for the real and imaginary parts). This is intended to emulate perfect CSI at the TX. 
 
    \item \emph{Quantized feedback}: Let $\hat{\nbh}_i^{\rm Q} = [\hat{h}_{i1}^{\rm Q} ~ \hat{h}_{i2}^{\rm Q}]$ denote the quantized version of $\hat{\nbh}_i$, where each complex $\hat{h}_{il}^{\rm Q} ~(i,l \in \{1,2\})$ is an 8-bit representation of $\hat{h}_{il}$  (i.e., 4 bits for the real and imaginary parts). The quantization mechanism is described in Appendix~\ref{appendix:A}, and is based on based on IEEE 802.11n \cite{IEEE802.11n2009}. It results in a more than $90\%$ reduction in the CSI feedback overhead, and is intended to emulate imperfect CSI at the TX. 
\end{itemize}
The fedback CSI is used by the TX to design the precoders, as described in the following subsection.
}
 
\subsection{ Stage 2: RSMA Signal Design}
\label{subsec:stage 2}

\begin{figure*}[ht]
\centering 
\includegraphics[width=0.99\textwidth]{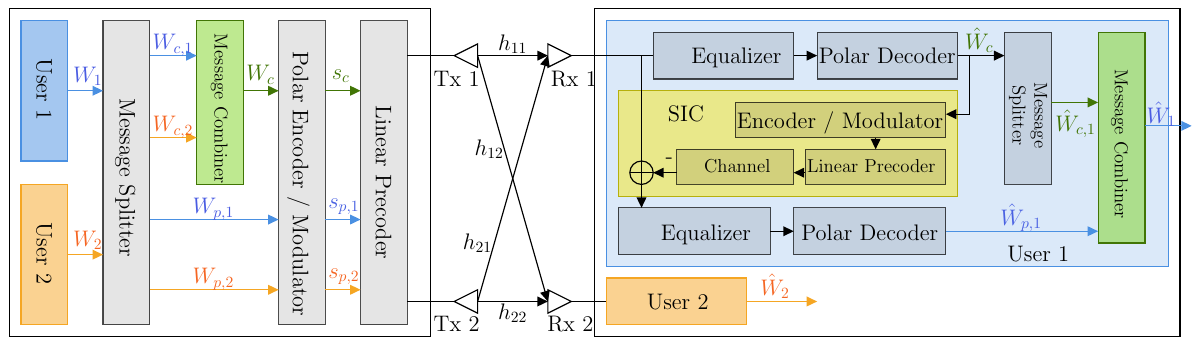}
  \caption{RSMA block diagram for the two-user MISO scenario.}
  \label{fig:rsma 2 user diagram}
\end{figure*}
    For the two-user MISO scenario, Fig. \ref{fig:rsma 2 user diagram} depicts RSMA in operation. Let $W_1$ and $W_2$ denote the messages corresponding to RX~1 and RX~2, respectively. At the TX, each $W_i~(i=1,2)$ is split into common and private portions -- denoted by $W_{c,i}$ and $W_{p,i}$, respectively -- by the message splitter. The common portions of each RX's message {(i.e., $W_{c,1}$ and $W_{c,2}$)} are combined into a common message, which is then encoded { and modulated} to form a common stream, {$s_c[k]~(k = 0,\cdots,N_c-1)$ over the subcarriers}. On the other hand, the private portions of RX~$i$'s message are individually encoded { and modulated} to form private streams, {$s_i[k]$. All three streams (i.e., $s_c[k], s_1[k]$ and $s_2[k]$) are linearly precoded to form the transmit OFDM frequency domain signal as follows:
    \begin{equation}
        \label{eq:transmit stream equation}
        \mathbf{x}[k] = \mathbf{p}_c s_c[k] + \mathbf{p}_1 s_1[k] + \mathbf{p}_2 s_2[k]~ (k = 0, \cdots, N_c - 1),
    \end{equation}
    where $\nbp_c$ is referred to as the common stream precoder, and $\nbp_i$ the private stream precoder of RX~$i$.     
  
    After appropriate DFT-based processing, the received OFDM frequency domain signal, $y_i[k]$, at RX~$i$ is given by:
    \begin{align}
        \label{eq:RX k received signal}
        y_i[k] &= \mathbf{h}_i^H[k] \mathbf{x}[k] + n_i [k], ~ (k = 0, \cdots, N_c-1) \notag \\
            &= \nbh_i^H[k] \nbp_c s_c [k] + \nbh_i^H[k] \nbp_1 s_1 [k] +  \nbh_i^H[k] \nbp_2 s_2[k] \notag \\
            &~ + n_i[k] ~ (i=1,2),
    \end{align}
    where $n_i[k] \sim \mathcal{CN}(0,\sigma^2)$ is the thermal noise. We assume that precoded pilot signals have been interspersed in some of the subcarriers of $\nbx[k]$ to help RX~$i$ estimate $\nbh_i^{H}[k] \nbp_c$ and $\nbh_i^{H}[k] \nbp_i$ upon receiving $y_i[k]$\footnote{These pilots play a role similar to the demodulation reference signals (DM-RS) used in LTE and 5G NR.}. Then, after suitable equalization (e.g., MMSE), RX~$i$ first decodes $s_c[k]$ to recover $W_c$, while treating the interference from the $s_1[k]$ and $s_2[k]$ as noise. The decoded estimate of $W_c$ at RX~$i$, denoted by $\hat{W}_c^{(i)}$, is simultaneously sent to:
    \begin{itemize}
        \item[(a)] the message splitter to extract  $\hat{W}_{c,i}^{(i)}$ -- the portion of $\hat{W}_c^{(i)}$ containing data meant for RX~$i$'s, and

        \item[(b)] the SIC module, which generates an estimate of $\nbh_i^H[k] \nbp_c s_c[k]$ and subtracts it from $y_i[k]$. The residue is once again subjected to equalization, this time to decode $s_i[k]$ by treating the interference from the other private stream as noise. Let $\hat{W}_{p,i}$ denote the decoded estimate of $W_{p,i}$ at RX~$i$. 
    \end{itemize} 

    Let $\nbP = [\nbp_c ~ \nbp_1 ~ \nbp_2]$. As a function of $\nbP$, from (\ref{eq:RX k received signal}), the achievable rate (in bits/s/Hz) for error-free decoding of $s_c[k]$ at RX~$i$, denoted by $R_{c,i}(\nbP)$, is given by:
    \begin{align}
    \label{eq:achievable common rate}
        R_{c,i}(\nbP) &= \min_k \log_2 \left(1 + \frac{|\mathbf{h}_i^H[k] \mathbf{p}_c|^2}{\sigma^2 + |\mathbf{h}^H_i[k] \mathbf{p}_1|^2 + |\mathbf{h}^H_i [k] \mathbf{p}_2|^2} \right),
    \end{align}
    where the minimum over $k$ arises because the encoded and modulated common stream is embedded over subcarriers in the frequency domain. Additionally, since both RXs must first decode $s_c[k]$, it follows that the achievable rate for error-free decoding of $s_c[k]$ at \emph{both} RXs, denoted by $R_c(\nbP)$, is given by:
    \begin{align}
    \label{eq:total common rate}
        R_c(\nbP) &= \min (R_{c,1}(\nbP), R_{c,2}(\nbP)).
    \end{align}
Assuming perfect SIC operation, the achievable rate for error-free decoding of $s_i[k]$ at RX~$i$, similar to (\ref{eq:achievable common rate}), is given by:
    \begin{align}
    \label{eq:achievable private rate}
    R_i(\nbP) = \min_k \log_2 \left( 1+\frac{|\mathbf{h}_i^H[k] \mathbf{p}_i|^2}{\sigma^2 + |\mathbf{h}_i^H[k] \mathbf{p}_j|^2} \right),~ j\neq i.
    \end{align}
Thus, the achievable rate at RX~$i$ for error-free recovery of $W_i$, denoted by $R_i^{\rm RSMA}(\nbP)$ is given by:
    \begin{align}
    \label{eq:maxSE_RXi}
        R_i^{\rm RSMA}(\nbP) &= (W_{c,i}/ W_c) R_c(\nbP) + R_i(\nbP)
    \end{align}
where the first term captures the fraction of the common message, $W_c$, intended for RX~$i$. Consequently, the achievable RSMA sum rate, denoted by $R^{\rm RSMA}(\nbP)$, is:
    \begin{align}
        \label{eq:maxSE_sum}
        R^{\rm RSMA}(\nbP) &= R_1^{\rm RSMA}(\nbP) + R_2^{\rm RSMA}(\nbP) \notag \\
                    &= R_c(\nbP) + R_1(\nbP) + R_2(\nbP)
    \end{align}
The canonical \emph{sum rate maximization problem}, where the precoders at the TX are optimized to maximize $R^{\rm RSMA}(\nbP)$ is formulated as follows:
    \begin{align}
    \label{prob: OP_ideal}
    OP_{\rm sr}^{\rm RSMA}: \max_{\nbP} &~ R^{\rm RSMA}(\nbP) \\
    \label{const: OP_ideal_tx_power}
    \mbox{s.t.} &~ {\rm tr}(\nbP \nbP^H) \leq P_t,
    \end{align}
where (\ref{const: OP_ideal_tx_power}) restricts the transmit power to $P_t$. Through the well-known Rate-WMMSE relationship \cite{Christensen_etal_2008}, the non-convex problem in $OP_{\rm sr}^{\rm RSMA}$ can be transformed into a sequence of tractable sub-problems that can be iteratively solved till convergence to a local optimum. This is commonly known in the literature as the \emph{WMMSE method} and despite only guaranteeing local optimality, it is widely recognized as a benchmark solution for the family of MIMO precoder design problems, of which $OP_{\rm sr}^{\rm RSMA}$ is a specific instance corresponding to RSMA. Let $\nbP_{\rm wmmse}^{\rm RSMA}$ denote the precoders obtained by applying the WMMSE method to solve $OP_{\rm sr}^{\rm RSMA}$.
\begin{nrem}[Imperfect CSI at the TX]
\label{rem:CSIT}
  $OP_{\rm sr}^{\rm RSMA}$ contains a slight abuse of notation for the sake of simplicity. Let $\hat{\nbH}^{\rm TX} := [\hat{\nbh}_1^{\rm TX} ~ \hat{\nbh}_2^{\rm TX}]$ denote the CSIT, where $\hat{\nbh}_i^{\rm TX} \in \{\hat{\nbh}_i^{\rm UQ}, \hat{\nbh}_i^{\rm Q}\}$, as described in Section~\ref{subsec:stage1}. Strictly speaking, to obtain the objective function of $OP_{\rm sr}^{\rm RSMA}$ \underline{at the TX}, $\nbh_i[k]$ in (\ref{eq:achievable common rate}) and (\ref{eq:achievable private rate}) should be replaced by $\hat{\nbh}_i^{\rm TX}$, so that the solution to $OP_{\rm sr}^{\rm RSMA}$ becomes a function of the CSIT.
\end{nrem}

In this paper, the main performance metric in our measurements is the throughput (measured in bits/s). Over an effective bandwidth\footnote{The effective bandwidth is the portion of the total bandwidth available for data transfer, after accounting for signaling overheads (e.g., cyclic prefix in the case of OFDM).}, $B$, the achievable RSMA sum throughput, denoted by $T^{\rm RSMA}(\nbP)$, is related to $R^{\rm RSMA}(\nbP)$ in (\ref{eq:maxSE_sum}) via $T^{\rm RSMA}(\nbP) = B R^{\rm RSMA}(\nbP)$. Hence, it follows that $T^{\rm RSMA}(\nbP^{\rm RSMA}_{\rm wmmse})$ is a good benchmark in terms of the maximum achievable RSMA sum throughput. However, this benchmark is hard to realize in practice because:
\begin{itemize}
    \item[1.] In state-of-the-art communications systems, PHY layer transmissions are restricted to a finite collection of modulation and coding scheme (MCS) levels (e.g., QPSK, rate $1/2$), which caps the maximum achievable sum throughput, and
        
    \item[2.] Error-free decoding of the transmitted messages at the RX cannot be guaranteed due to finite block length effects, which further diminishes the realized throughput.
\end{itemize} 
We address these two points by first characterizing the \emph{MCS-limited} throughput. An MCS level can be defined by a pair $(m, r)$, where positive integer $m$ denotes the bits per constellation symbol (e.g., 2 for QPSK) and $r \in (0,1]$ denotes the code rate. Hence, the achievable rate (in bits/s/Hz) for MCS level $(m,r)$ equals $mr$, and the achievable throughput is $Bmr$. Let $(m_c, r_c)$, $(m_1, r_1)$ and $(m_2, r_2)$ denote the MCS levels chosen for $s_c[k]$, $s_1[k]$ and $s_2[k]$, respectively. Then, the \emph{MCS-limited} RSMA sum throughput, denoted by $T^{\rm RSMA}_{\rm mcs}(\nbP, \ncalM)$, is given by: 
\begin{align}
    \label{eq:mcstput_sum}
    T^{\rm RSMA}_{\rm mcs}(\nbP, \ncalM) &= B m_c r_c(\nbp_c) \times \nbbP(\hat{W}_c^{(1)} = W_c; \hat{W}_c^{(2)} = W_c) \notag \\
    &~ + B m_1 r_1(\nbp_1) \times \nbbP(\hat{W}_{p,1} = W_{p,1}) \notag \\
    &~ + B m_2 r_2(\nbp_2) \times \nbbP(\hat{W}_{p,2} = W_{p,2}).
\end{align}
The RHS of (\ref{eq:mcstput_sum}) indicates that the MCS levels of the streams are, in general, a function of the corresponding precoders. The probability terms capture the loss in throughput due to decoding errors -- specifically, $\nbbP(\hat{W}_c^{(1)} = W_c; \hat{W}_c^{(2)} = W_c)$ denotes the probability that $W_c$, is correctly decoded at both RXs, while $\nbbP(\hat{W}_{p,i} = W_{p,i})$ denotes the probability that $W_{p,i}$ is correctly decoded at RX~$i$. Thus, similar to $OP_{\rm sr}^{\rm RSMA}$ in (\ref{prob: OP_ideal})-(\ref{const: OP_ideal_tx_power}), the \emph{MCS-limited RSMA sum throughput maximization problem}, which is of considerable practical relevance, can be defined as follows:
\begin{align}
    \label{prob: OP_meas}
    OP_{\rm mcs}^{\rm RSMA}: 
    \max_{\nbP, \ncalM} &~ T_{\rm mcs}^{\rm RSMA}(\nbP^{\rm RSMA}_{\rm wmmse}, \ncalM) \\
    \label{eq:P0_mcs}
    &~ \ncalM \in \nbbM,
\end{align}
where $\nbbM$ denotes the collection of permissible MCS levels for the three streams, which is typically pre-determined through standards (see Table~\ref{tab: Mcs table} for the $\nbbM$ used in our measurements). $OP_{\rm mcs}^{\rm RSMA}$ does not have a closed form solution, as the message decoding probabilities in (\ref{eq:mcstput_sum}) are difficult to characterize in terms of $\nbP_{\rm wmmse}^{\rm RSMA}$ and $\ncalM$. However, these probabilities can be empirically evaluated through measurements. This motivates us to empirically solve $OP_{\rm mcs}^{\rm RSMA}$ through a brute force search\footnote{A more sophisticated empirical approach involves link adaptation, where the most suitable MCS level is determined by an ARQ-based mechanism \cite{mosquera2023linkadaption}. This is left for future work.} over $\nbbM$, which is the focus of our measurements in Section~\ref{sec: measurement results} using our SDR-based RSMA prototype. We conclude this subsection with the following remark:

\begin{nrem}[Wideband precoders]
    In principle, narrowband precoders for each subcarrier (i.e., $\nbp_c[k]$, $\nbp_1[k]$, $\nbp_2[k]$: $k = 0, \cdots, N_c - 1$) could increase the achievable rate in (\ref{eq:achievable common rate})-(\ref{eq:maxSE_sum}) by adapting to channel variations at subcarrier-level granularity. However, this would involve huge signalling overhead in both directions, as: (a) narrowband CSI (i.e., $\hat{\nbh}_i[k]$ for several, if not all, $k$) would need to be fed back to the TX for precoder design, and (b) more demodulation reference signals would have to be embedded in the transmit signal to estimate $\nbh_i^H[k]\nbp_c[k]$ and $\nbh_i^H[k]\nbp_i[k]$ at RX~$i$ for all $k$. Thus, to reduce the signalling overhead, we assume the same precoders, $\nbP$, across all the subcarriers.
\end{nrem}
}

\subsection{ RSMA v/s SDMA v/s NOMA}
{In general, the splitting of the user messages in Fig.~\ref{fig:rsma 2 user diagram} need not be equal (i.e., 50-50) between the common and private portions, nor does the ratio between the common and private portions need to be fixed across both messages. In fact, SDMA and NOMA are special cases of RSMA corresponding to specific choices of message splitting, as explained in the following remarks.} 
\begin{nrem}[SDMA as a special case of RSMA]
\label{rem:rsma_spl_case_sdma}
For the special case where the common stream is turned off  -- i.e., $W_i = W_{p,i}~(i = 1,2)$ in Fig.~\ref{fig:rsma 2 user diagram} -- RSMA reduces to SDMA \cite{RSMAUnifying}.
\end{nrem}

{
\begin{nrem}[NOMA as a special case of RSMA]
\label{rem:rsma_spl_case_noma}
Without loss of generality, suppose RX~2 has a weaker channel than RX~1. For the special case where the private stream of RX~2 is turned off, and no portion of $W_1$ is allocated to the common stream -- i.e., $W_1 = W_{p,1}, W_2 = W_{c,2} = W_c$ in Fig.~\ref{fig:rsma 2 user diagram} --  RSMA reduces to NOMA \cite{RSMAUnifying}.
\end{nrem}
This flexibility in message splitting provided by RSMA characterizes its interference management philosophy, as follows:
\begin{itemize}
    \item \textbf{SDMA}: In the absence of a common stream, the interference experienced at RX~$i$ due to $W_{j}~(j \neq i)$ is treated as noise.
    \item \textbf{NOMA}: The common stream allows the stronger RX~1 to fully decode and subtract the interference it experiences (from $W_2$). The weaker RX~2 continues to treat the interference it experiences (from $W_1$) as noise, like SDMA.
    \item \textbf{RSMA}: The interference at RX~$i$ due to $W_{j}~(j \neq i)$ is \emph{partially decoded} ($W_{c,j}$) and \emph{partially treated as noise} ($W_{p,j}$). This feature enables RSMA to softly bridge between SDMA and NOMA \cite{Mao2018, RSMAUnifying}. In other words, when the two channels are orthogonal, the optimal solution to $OP_{\rm sr}^{\rm RSMA}$ in (\ref{prob: OP_ideal})-(\ref{const: OP_ideal_tx_power}) coincides with SDMA. Similarly, when the channels are aligned and one of them is much stronger than the other, the optimal solution to $OP_{\rm sr}^{\rm RSMA}$ reduces to NOMA. Importantly, over the vast majority of channels that do not fall in either extreme, RSMA is expected to provide significant gains over both SDMA and NOMA, which we aim to demonstrate experimentally in Section~\ref{sec: measurement results}.
\end{itemize}

Based on Remarks~\ref{rem:rsma_spl_case_sdma} and \ref{rem:rsma_spl_case_noma}, the expressions corresponding to SDMA and NOMA in (\ref{eq:transmit stream equation})-(\ref{eq:P0_mcs}) can be obtained by discarding $\nbp_c$ (SDMA) and $\nbp_2$ (NOMA). These expressions are provided in Table~\ref{tab:rsma_noma_sdma} for clarity. In Sections~\ref{subsec:sum_tput_comparison} and \ref{subsec:fairness_comparison}, we provide an experimental RSMA v/s SDMA v/s NOMA performance comparison by empirically solving $OP_{\rm mcs}^{\rm ma},~ {\rm ma} \in \{{\rm RSMA, SDMA, NOMA}\}$ using our SDR-based RSMA prototype. In the following section, we present details of our prototype, as well as how the system model described above in Sections~\ref{subsec:stage1} and \ref{subsec:stage 2} is implemented.
}
{
\begin{table*}[ht]
    \centering
    {
    \begin{tabular}{|l|l@{}r@{}l|} 
      \hline 
        Quantity &  &{} Expression &{} \\
      \hline
       Message   &       &{} $W_i$ &{} $ = W_{c,i} \cup W_{p,i}~ (i = 1,2)$\\
       splitting &       &{} $W_c$ &{} $ = W_{c,1} \cup W_{c,2}$ \\
                 & RSMA: &{} $W_{c,i}, W_{p,i}$ &{} $\neq \emptyset $, in general  \\ 
                 & SDMA: &{} $W_{c,i}$ &{} $= \emptyset, W_{p,i} = W_i$ \\
                 & NOMA: &{} $W_{c,1}$ &{} $= \emptyset, W_{p,1} = W_1$ and $W_{c,2} = W_2, W_{p,2} = \emptyset$ \\
      \hline
      Precoders & RSMA: &{} $\nbP$ &{} $= [\nbp_c ~ \nbp_1 ~ \nbp_2]$ \\
                & SDMA: &{} $\nbP$ &{} $= [\nbp_1 ~ \nbp_2]$ \\
                & NOMA: &{} $\nbP$ &{} $= [\nbp_c ~ \nbp_1]$ \\
      \hline
      Achievable & RSMA: &{} $R_{c,i}(\nbP)$ &{} $= \displaystyle\min\limits_k \log_2 \left(1 + \frac{|\mathbf{h}_i^H[k] \mathbf{p}_c|^2}{\sigma^2 + |\mathbf{h}^H_i[k] \mathbf{p}_1|^2 + |\mathbf{h}^H_i [k] \mathbf{p}_2|^2} \right)$ \\
      common stream           &       &{} $R_c(\nbP)$ &{} $= \displaystyle\min\limits_i R_{c,i}(\nbP)$ \\
      rate & SDMA: &{} $R_{c,i}(\nbP)$ &{} $= R_c(\nbP) = 0$ \\
      at RX~$i$ & NOMA: &{} $R_{c,i}(\nbP)$ &{} $= \displaystyle\min\limits_k \log_2 \left(1 + \frac{|\mathbf{h}_i^H[k] \mathbf{p}_c|^2}{\sigma^2 + |\mathbf{h}^H_i[k] \mathbf{p}_1|^2 } \right)$ \\
                 &       &{} $R_c(\nbP)$ &{} $ = \displaystyle\min\limits_i R_{c,i}(\nbP)$ \\ 
    \hline
      
    Achievable  pvt. &  RSMA: &{} $R_i(\nbP)$ &{} $= \displaystyle\min\limits_k \log_2 \left(1 + \frac{|\mathbf{h}_i^H[k] \mathbf{p}_i|^2}{\sigma^2 + |\mathbf{h}^H_i[k] \mathbf{p}_j|^2} \right)~ (j \neq i)$ \\
    stream rate  &  SDMA: &{} $R_i(\nbP)$ &{} $= \displaystyle\min\limits_k \log_2 \left(1 + \frac{|\mathbf{h}_i^H[k] \mathbf{p}_i|^2}{\sigma^2 + |\mathbf{h}^H_i[k] \mathbf{p}_j|^2} \right)~ (j \neq i)$ \\
    at RX $i$ & NOMA: &{} $R_1(\nbP)$ &{} $ =  \displaystyle\min\limits_k \log \left(1 + \frac{|\mathbf{h}_1^H[k] \mathbf{p}_1|^2}{\sigma^2 } \right)$ \\
                    &       &{} $R_2(\nbP)$ &{} $ = 0$ \\
      \hline
      Achievable & RSMA: &{} $R^{\rm RSMA}(\nbP)$ &{} $= R_c(\nbP) + R_1(\nbP) + R_2(\nbP)$ \\
      sum rate & SDMA: &{} $R^{\rm SDMA}(\nbP)$ &{} $= R_1(\nbP) + R_2(\nbP)$ \\
      & NOMA: &{} $R^{\rm NOMA}(\nbP)$ &{} $= R_c(\nbP) + R_1(\nbP)$ \\
      \hline
      Sum rate   & &{} $OP_{\rm sr}^{\rm ma}:~\displaystyle\max\limits_\nbP$ &{} $R^{\rm ma}(\nbP) $ \\
      max.~problem   & &{} $\mbox{s.t.}$ &{} ${\rm tr}(\nbP \nbP^H) \leq P_t$ \\ 
      \hline
      Achievable & &{} $T^{\rm ma}(\nbP)$ &{} $= B R^{\rm ma}(\nbP)$,\\
      sum throughput &   &{} where ${\rm ma}$ &{} $\in \{\rm{RSMA, SDMA, NOMA}\}$  \\
      \hline
      MCS levels & RSMA: &{} $\ncalM$ &{} $= \{(m_c, r_c), (m_1, r_1), (m_2, r_2) \}$ \\
                 & SDMA: &{} $\ncalM$ &{} $= \{(m_1, r_1), (m_2, r_2) \}$ \\
                 & NOMA: &{} $\ncalM$ &{} $= \{(m_c, r_c), (m_1, r_1)\}$ \\
      \hline
      MCS-limited & RSMA: &{} $R_{\rm mcs}^{\rm RSMA}(\nbP, \ncalM)$ &{} $ = m_c r_c(\nbp_c) + m_1 r_1 (\nbp_1) + m_2 r_2 (\nbp_2)$ \\
      sum rate & SDMA: &{} $R_{\rm mcs}^{\rm SDMA}(\nbP, \ncalM)$ &{} $ = m_1 r_1 (\nbp_1) + m_2 r_2 (\nbp_2)$ \\
       & NOMA: &{} $R_{\rm mcs}^{\rm NOMA}(\nbP, \ncalM)$ &{} $ = m_c r_c(\nbp_c) + m_1 r_1 (\nbp_1)$ \\
      \hline
      MCS-limited & RSMA: &{} $T^{\rm RSMA}_{\rm mcs}(\nbP, \ncalM)$ &{} $= Bm_cr_c(\nbp_c) \times \nbbP(\hat{W}_c^{(1)} = W_c; \hat{W}_c^{(2)} = W_c)$\\
      sum throughput & &{} &{} ~$+ Bm_1r_1(\nbp_1) \times \nbbP(\hat{W}_{p,1} = W_{p,1})$ \\
      & &{} &{} ~$+ Bm_2r_2(\nbp_2) \times \nbbP(\hat{W}_{p,2} = W_{p,2})$\\
       & SDMA: &{} $T^{\rm SDMA}_{\rm mcs}(\nbP, \ncalM)$ &{} $= Bm_1r_1(\nbp_1) \times \nbbP(\hat{W}_{p,1} = W_1)$ \\
       & &{} &{} ~ $+ Bm_2r_2(\nbp_2) \times \nbbP(\hat{W}_{p,2} = W_2)$\\
      & NOMA: &{} $T^{\rm NOMA}_{\rm mcs}(\nbP, \ncalM)$ &{} $= Bm_cr_c(\nbp_c) \times \nbbP(\hat{W}_c^{(1)} = W_2; \hat{W}_c^{(2)} = W_2)$ \\
      & &{} &{} ~ $+ Bm_1r_1(\nbp_1) \times \nbbP(\hat{W}_{p,1} = W_1)$ \\
      \hline
      MCS-limited & &{} $OP_{\rm mcs}^{\rm ma}:~ \displaystyle\max\limits_{\ncalM}$ &{} $T^{\rm ma}_{\rm mcs}(\nbP^{\rm ma}_{\rm wmmse}, \ncalM)$\\
      sum throughput     & &{} $\mbox{s.t.}$ &{} $\ncalM \in \nbbM$ \\
      max. problem   &  &{} &{}   \\ 
      (Experiments & &{} &{} \\
      in this paper)   & &{} &{} \\ 
      \hline
    \end{tabular}
    }
    \caption{System model differences between RSMA, SDMA and NOMA w.r.t the terminology in Section~\ref{subsec:stage 2}.}
    \label{tab:rsma_noma_sdma}
\end{table*}
}
 
\section{RSMA Prototype using SDR}
\label{sec: RSMA SDR prototype}
\subsection{Hardware setup}
\label{sec:overall set up}
\begin{figure*}[ht]
    \centering
    \includegraphics[width=0.9\textwidth]{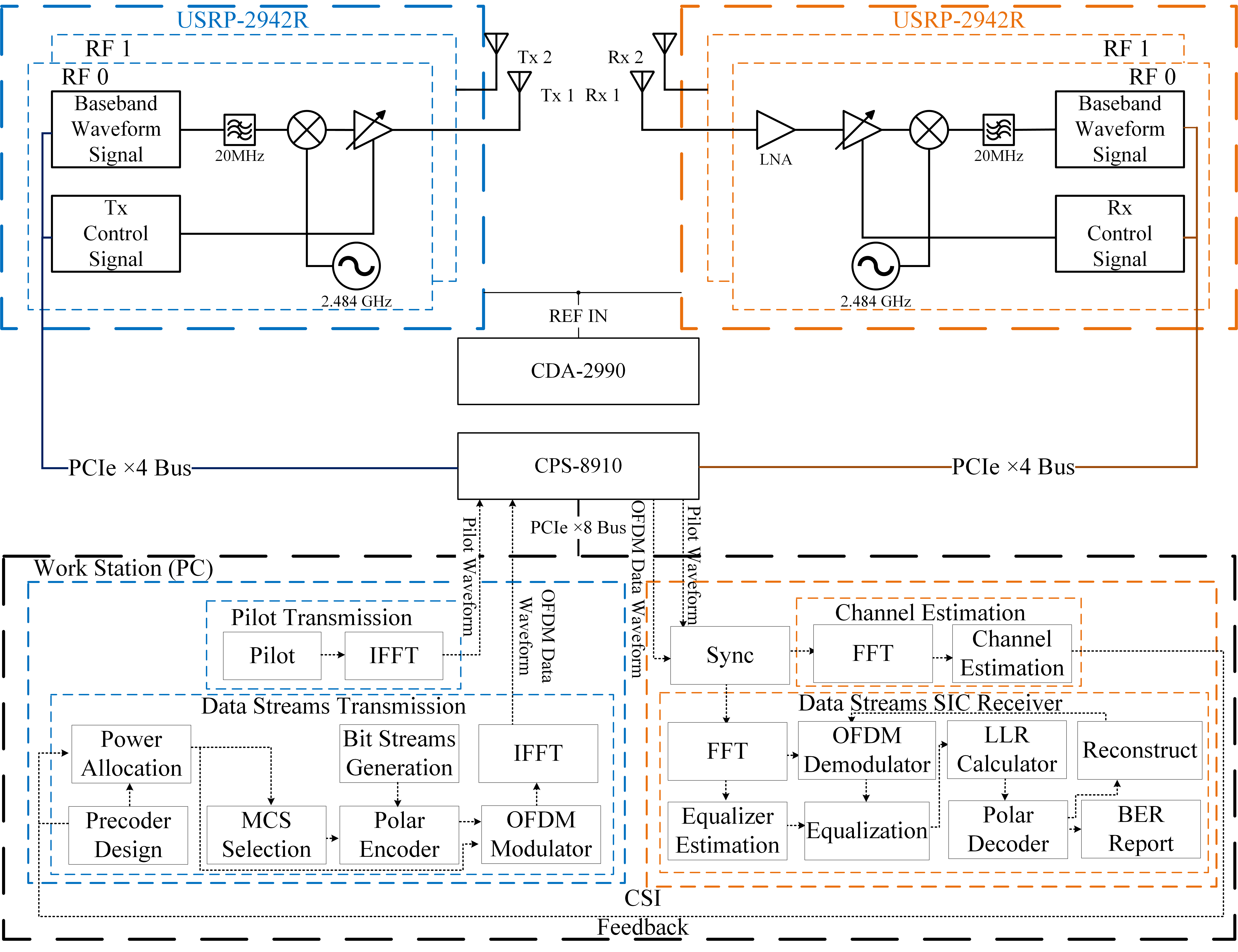}
    \caption{Block diagram of RSMA Prototype.}
    \label{fig:RSMA prototype Diagram Hardware}
\end{figure*}
Our RSMA prototype is built using National Instruments' (NI) Universal Software Radio Peripherals (USRPs). As shown in Fig. \ref{fig:RSMA prototype Diagram Hardware}, two USRP-2942 units are used to realize the two-user MISO scenario. In particular, the RX antennas (TP-LINK TL-ANT2405C) are connected to its USRP using coaxial cables, which allows them to be moved to create different channel configurations (Fig.~\ref{fig:all the cases in the measurement}). The USRPs share a common timing source (CDA-2990), and are connected to a workstation through which they are controlled using LabVIEW NXG. All connections are through PCIe cables, facilitated by a PCIe bus (CPS-8910). The total transmit power, $P_t$, is $23{\rm dBm}$. A list of hardware components is provided in Table \ref{tab: hardware in RSMA}. 

\begin{table*}
\centering
\begin{tabular}{|c|c|c|ll}
\cline{1-3}
  \textbf{No.} & \textbf{Name} & \textbf{Units Description}&  \\ \cline{1-3}
 1. & Workstation  & Running LabVIEW NXG     \\ 
 2. & NI USRP-2942 & SDRs used to realize TX and RXs    \\
 3. & NI CPS-8910  & Provides additional PCIe ports\\
 4. & NI CDA-2990 & 8 Channel, 10 MHz Clock Distribution Device \\ 
 5. & Linx Technologies ANT-W63WSx &  TX antennas  \\
 6. & TP-LINK TL-ANT2405C & RX antennas  \\ \cline{1-3}
\end{tabular}
\caption{List of hardware components.}
\label{tab: hardware in RSMA}
\end{table*}

\subsection{RSMA Implementation Details}
\label{subsec:challenges}
 \begin{figure*}
     \centering
     \includegraphics[width = 0.8\textwidth]{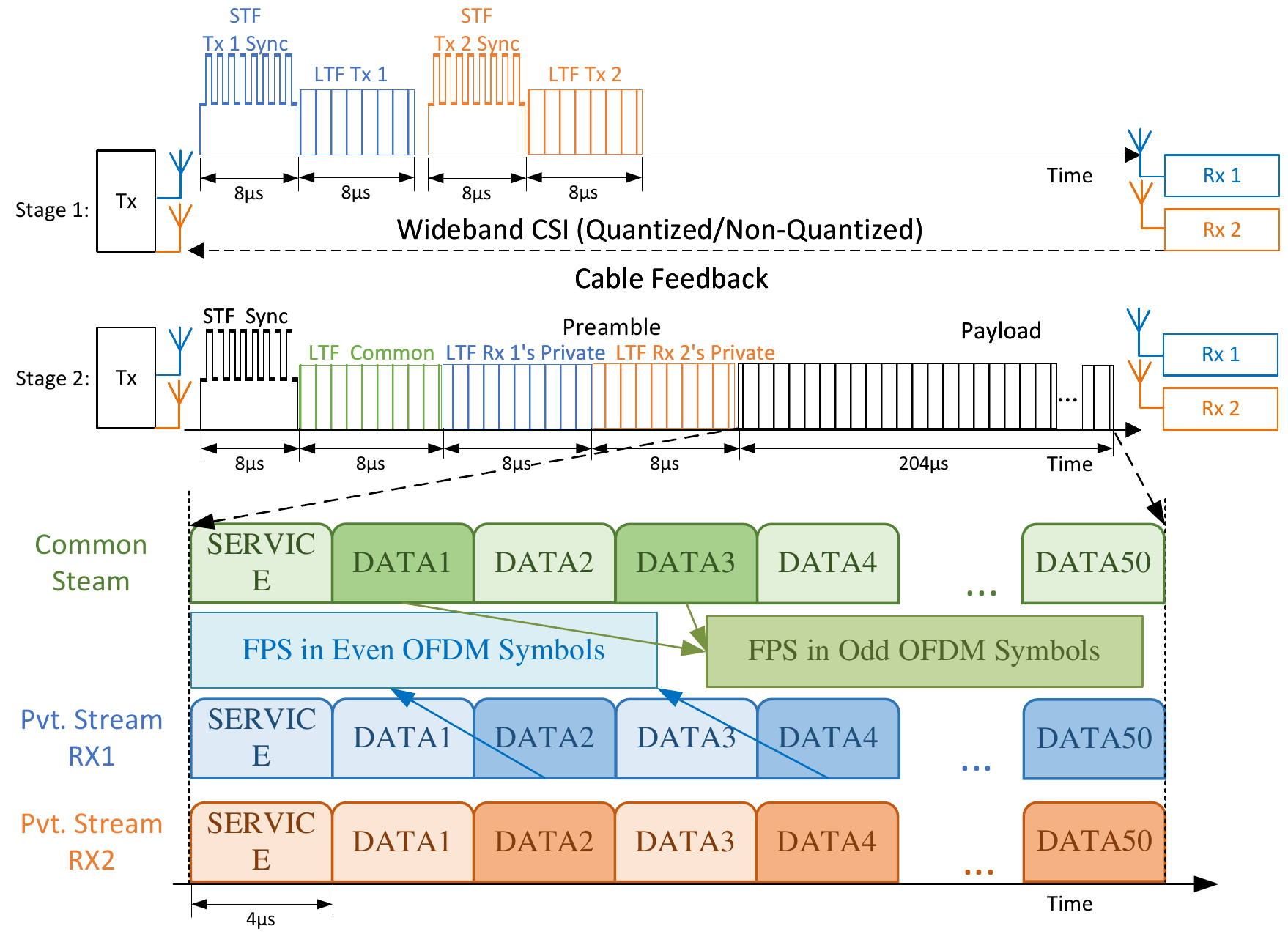}
     \caption{(Top and middle): The two-stage transmission protocol; (Bottom): OFDM symbol structure within the payload. In conventional 802.11g, every DATA symbol for an RX contains four pilot subcarriers for FPS. However, for RSMA, this approach would cause the pilot subcarriers in the common and private streams to interfere with each other. In particular, the inability to do FPS effectively for the common stream leads to large phase noise. To avoid this, the common (private) stream(s) carry non-zero pilots only in the odd (even) DATA symbols \cite{Framestructure}.}
     \label{fig:Overall diagram of SDR platform}
 \end{figure*}
The USRPs are capable of operating in the $2.4{\rm GHz}$ ISM band. Hence, we choose a center frequency of $2.484{\rm GHz}$ for our transmissions\footnote{This corresponds to channel no. $14$ in the IEEE 802.11 family of standards for the $2.4${\rm GHz} band. We use this channel to avoid ambient WiFi interference, as it is not commercially used in the UK.}. For ease of practicalities like synchronization and channel estimation, it is convenient to implement signals based on well-known physical layer standards (e.g., IEEE 802.11 frames) in a prototype. Hence, we adopt several features of the IEEE 802.11g physical layer frames in our signal design. {Next, we describe in detail how Stages 1 and 2 from Sections~\ref{subsec:stage1} and \ref{subsec:stage 2}, respectively, are implemented in our prototype: 

\paragraph{Stage 1}
Each TX antenna transmits a pilot signal orthogonally in time comprising a Short Training Field (STF, $8\mu{\rm s}$ in duration) and a Long Training Field (LTF, $8\mu{\rm s}$ in duration), as shown in the top portion of Fig.~\ref{fig:Overall diagram of SDR platform}. The STF is used for synchronization and coarse frequency offset estimation, while the LTF is used to obtain a least squares estimate of the CSI at each RX over all the subcarriers (i.e., $\hat{\nbh}_i[k]$ in Section~\ref{subsec:stage1}).

\paragraph{Stage 2}
Here, the transmitted signal consists of a preamble followed by the data payload, as shown in the middle portion of Fig.~\ref{fig:Overall diagram of SDR platform}.  
    \begin{itemize}
        \item \textbf{Preamble}: The preamble consists of one STF and three LTFs. The function of the STF is the same as the first stage, while the LTFs are precoded in order to estimate the \emph{precoded CSI} for equalization at the RX; in particular, the first LTF is used to estimate $\nbh_i^H \nbp_c$ at RX~$i~(=1,2)$ for decoding the common stream, the second to estimate $\nbh_1^H \nbp_1$ to decode the private stream meant for RX~1, and the third to estimate $\nbh_2^H \nbp_2$ to decode the private stream meant for RX~2.
        
        \item \textbf{Data Payload}: For the payload, we use OFDM signals over a total bandwidth of $20{\rm MHz}$ with $N_c = 64$ subcarriers (labeled DATA in Fig.~\ref{fig:Overall diagram of SDR platform}) and a cyclic prefix (CP) of 16 samples per OFDM symbol. Aligned with IEEE 802.11 frames, $52$ subcarriers are used for communications while the rest serve as guard bands. Among these $52$ subcarriers, $48$ are used to carry data symbols, with the remaining used for fine phase shifting (FPS)\footnote{FPS is used to correct the common phase error across all subcarriers in one OFDM symbol \cite{FinephaseShifting}.}. This yields an effective bandwidth of:
        \begin{align}
        \label{eq:Beff}
            B = 20{\rm MHz} \times \underbrace{(64/80)}_\text{CP overhead} \times \underbrace{(48/64)}_{\substack{\text{Guard band} \\ \text{overhead}}} = 12{\rm MHz} 
        \end{align}
        The payload consists of three superposed streams (one common, two private), each comprising 50 OFDM symbols (labeled DATA 1 through DATA 50, as shown in the bottom portion of Fig.~\ref{fig:Overall diagram of SDR platform}).

        \item \textbf{MCS Implementation}: Table~\ref{tab: Mcs table} lists the MCS levels, $\nbbM$, implemented in our prototype. MCS indices 0 through 7 are identical to IEEE 802.11g, where 64-QAM is the largest supported constellation. To allow for the possibility of higher data rates due to RSMA's enhanced interference suppression, we implement two more MCS levels supporting 256-QAM (MCS indices 8 and 9). After the preamble, the first OFDM symbol (labelled SERVICE in the bottom portion of Fig.~\ref{fig:Overall diagram of SDR platform}) contains the MCS information of each stream. 
        
        Unlike the IEEE 802.11 family of standards, which uses LDPC codes for error correction, we implement Polar codes augmented with an 8-bit cyclic redundancy check \cite{trifonovPolar, constructionPolar}, along with successive cancellation list decoding \cite{listdecoding}, with a list depth of 2. 
        
        \begin{table*}
        \centering
        \begin{tabular}{|c|c|c|c|}
        \hline
        MCS Index   & Modulation {($m$)} & Code Rate, $r$ & Data Rate, $Bmr$ (Mbps) \\ 
        \hline
        $0$         & BPSK {(1)}    & $1/2$         & $6$               \\ \hline
        $1$         & BPSK {(1)}    & $3/4$         & $9$               \\ \hline
        $2$         & QPSK {(2)}    & $1/2$         & $12$              \\ \hline
        $3$         & QPSK {(2)}    & $3/4$         & $18$              \\ \hline
        $4$         & 16QAM {(4)}   & $1/2$         & $24$              \\ \hline
        $5$         & 16QAM {(4)}   & $3/4$         & $36$              \\ \hline
        $6$         & 64QAM {(6)}   & $2/3$         & $48$              \\ \hline
        $7$         & 64QAM {(6)}   & $3/4$         & $54$              \\ \hline
        $8$         & 256QAM {(8)}  & $3/4$         & $72$              \\ \hline
        $9$         & 256QAM {(8)}  & $5/6$         & $80$              \\ \hline
        \end{tabular}
        \caption{MCS levels (largely based on IEEE 802.11g) implemented in our prototype. The data rate in the last column is equal to $B m r$, where $B$ is the effective bandwidth given by (\ref{eq:Beff}).}
        \label{tab: Mcs table}
        \end{table*}
    \end{itemize}

An instance of Stage 1 and Stage 2, as described above and illustrated in Fig.~\ref{fig:Overall diagram of SDR platform}, constitutes a single measurement run. To empirically solve $OP_{\rm mcs}^{\rm ma} ~ ({\rm ma}\in \{{\rm RSMA, SDMA, NOMA}\})$ from Table~\ref{tab:rsma_noma_sdma}, we conduct $100$ measurement runs. Let $D_{s,c}^{\rm ma}$ denote the number of runs in which the common stream is successfully decoded by both RXs. Similarly, let $D_{s,i}^{\rm ma}~(i = 1,2)$ denote the number of runs in which RX~$i$ successfully decodes its private stream. Consider the expressions for the MCS-limited sum throughput in Table~\ref{tab:rsma_noma_sdma} (third last row). Replacing the decoding probabilities therein with their empirical estimates, the \emph{measured} sum throughput for each multiple access scheme is given by: 
        \begin{align}
            \label{eq: throughput calculation in RSMA SDR measurement}
            T_{\rm mcs}^{\rm RSMA}(\nbP_{\rm wmmse}^{\rm RSMA}, \ncalM) &= \frac{D_{s,c}^{\rm RSMA}}{100}B m_c r_c + \sum_{i=1}^2\frac{D_{s,i}^{\rm RSMA}}{100} B m_ir_i \\
            \label{eq: throughput calculation in SDMA SDR measurement}
            T_{\rm mcs}^{\rm SDMA}(\nbP_{\rm wmmse}^{\rm SDMA}, \ncalM) &=  \frac{D_{s,1}^{\rm SDMA}}{100} B m_1r_1 +  \frac{D_{s,2}^{\rm SDMA}}{100} B m_1r_1 \\
            \label{eq: throughput calculation in NOMA SDR measurement}
            T_{\rm mcs}^{\rm NOMA}(\nbP_{\rm wmmse}^{\rm NOMA}, \ncalM) &= \frac{D_{s,c}^{\rm NOMA}}{100}B m_c r_c +  \frac{D_{s,1}^{\rm NOMA}}{100} B m_1r_1, 
        \end{align}
where $B$ is given by (\ref{eq:Beff}) and the MCS levels are chosen from Table~\ref{tab: Mcs table}.
}

\section{Measurement Results}
\label{sec: measurement results}
{In this section, we first describe our measurement campaign in Section~\ref{subsec:measurement_campaign}. Then, in Section~\ref{subsec:sum_tput_comparison}, we present measurement results pertaining to $OP_{\rm mcs}^{\rm ma}$ with which we compare the measured sum throughput for RSMA, SDMA and NOMA. Then, in Section~\ref{subsec:fairness_comparison}, we present a fairness comparison between RSMA, SDMA and NOMA.}

\begin{figure*}
        \centering
        \includegraphics[width=\linewidth]{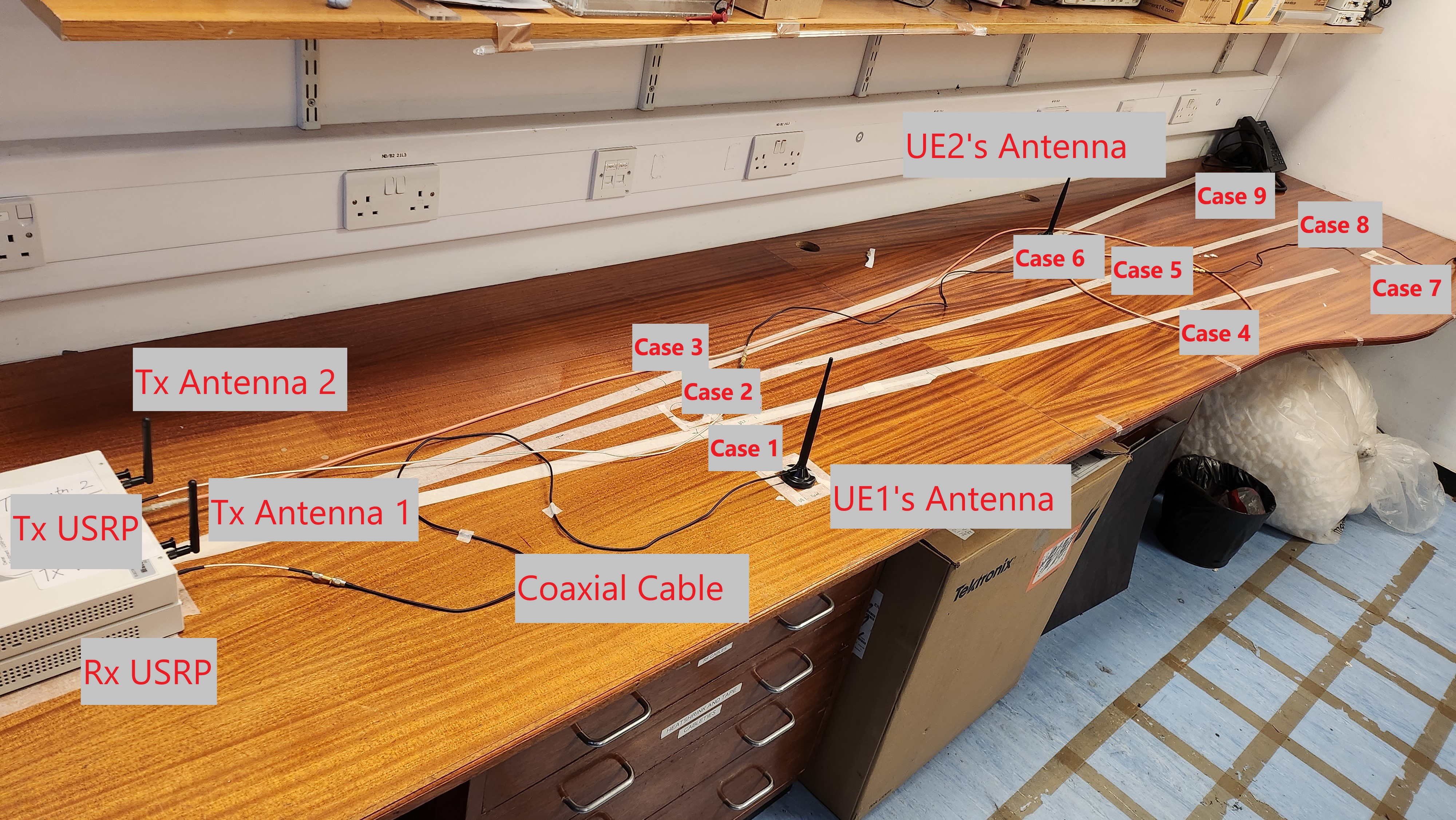}
        \caption{Layout of measurement cases. {The RX antenna positions in the above snapshot correspond to Case 6.}}
        \label{fig:all the cases in the measurement}
\end{figure*}

\subsection{Measurement campaign}
\label{subsec:measurement_campaign}
{Our measurements were carried out on a lab bench, as shown in Fig.~\ref{fig:all the cases in the measurement}}. The TX position is fixed along with the antenna of RX~1, which is placed $0.8{\rm m}$ away from the TX\footnote{As seen in Table~\ref{tab:param_list}, this is the closest TX-RX distance in our experiments. Since it exceeds the Fraunhofer distance, we are operating in the far-field.}. The antenna position of RX~2 is varied as marked in Fig.~\ref{fig:all the cases in the measurement} to create pairs of LoS channels with varying relative strengths and spatial correlation. {For NOMA in particular, RX~1 is treated as the strong user and RX~2 the weak user. Every attempt was made to avoid any relative motion between the TXs and RXs during the measurements. For easy reference, a full list of all the parameters relevant to our measurements is provided in Table~\ref{tab:param_list}.}
\begin{table}[]
    \centering
    {
    \begin{tabular}{|c|c|c|}
    \hline 
      Parameter & Notation (where applicable) & Value\\
    \hline
       Center frequency  & $f_c$  & $2.484{\rm GHz}$\\
       Transmit power & $P_t$ & $23{\rm dBm}$\\
       TX antenna length &  & $0.13{\rm m}$ \\
       Fraunhofer distance & & $0.28{\rm m}$\\
    \hline 
       Total bandwidth & & $20{\rm MHz}$ \\
       Subcarriers & Total ($N_c$) & $64$ \\
       & Data & $48$ \\
       & Pilot (FPS) & $4$\\
       & Guard band  & $12$ \\
       CP length & & $16$ \\
       Effective bandwidth & $B$ & $12{\rm MHz}$ \\ 
       OFDM symbols in payload & & $50$ \\
       Experiment runs (per case) & & $100$\\
       \hline       
       Distances & RX 1 $\rightarrow$ TX  & $0.8{\rm m}$\\
         & RX 2 $\rightarrow$ TX (Cases 1-3) & $1.25{\rm m}$\\
         & RX 2 $\rightarrow$ TX (Cases 4-6) & $2.10{\rm m}$\\
         & RX 2 $\rightarrow$ TX (cases 7-9) & $3.30{\rm m}$\\
    \hline
       Channel coding & & Polar codes \\
       List depth & & 2 \\
    \hline
       Environment & & LoS\\
    \hline
    \end{tabular}
    }
    \caption{List of parameters used in our experiments.}
    \label{tab:param_list}
\end{table}

{To capture the relative pathloss difference between the two channels, we introduce the channel strength disparity parameter, $\alpha$, defined in dB scale as follows:
\begin{align}
\label{eq:pathloss_param}
    \alpha~[{\rm dB}]= 10 \log_{10} \frac{\|\hat{\nbh}_2 \|}{\|\hat{\nbh}_1\|}.
\end{align}
where $\hat{\nbh}_i~(i = 1,2)$ is given by (\ref{eq:wideband_csir}). Hence, RX~2 has a weaker channel when $\alpha$ is negative.} 

Similarly, to measure the spatial correlation between the two channels, we introduce the correlation parameter, $\rho \in [0, 1]$, defined as follows:
\begin{align}
\label{eq:spatial_corr_param}
    \rho = 1 - \frac{|\hat{\nbh}_1^H \hat{\nbh}_2|^2}{\|\hat{\nbh}_1\| . \|\hat{\nbh}_2\|}.
\end{align}
Thus, $\rho = 0$ when the channels are completely aligned, whereas $\rho = 1$ when they are orthogonal. 


Based on (\ref{eq:pathloss_param}) and (\ref{eq:spatial_corr_param}), we make the following remarks on the cases marked in Fig.~\ref{fig:all the cases in the measurement}.

\begin{table}[]
    \centering
    {
    \begin{tabular}{|c|c|c||c|c|}
    \hline
    & \multicolumn{2}{|c||}{Relative pathloss} & \multicolumn{2}{|c|}{Spatial correlation} \\
    \hline
    & Cases & $\alpha$ [dB] & Cases & $\rho$ \\
    \hline
    Low & 1 & $-7.6$ & 1 & $0.15$ \\
        & 2 & $-6.4$ & 4 & $0.16$ \\
        & 3 & $-8.8$ & 7 & $0.24$ \\
    \hline
    Medium & 4 & $-14.5$ & 2 & $0.48$\\
           & 5 & $-13.0$ & 5 & $0.46$\\
           & 6 & $-13.0$ & 8 & $0.35$\\
    \hline
    High & 7 & $-23.6$ & 3 & $0.95$\\
         & 8 & $-24.7$ & 6 & $0.95$ \\
         & 9 & $-22.1$ & 9 & $0.85$ \\
    \hline
    \end{tabular}
    }
    \caption{The cases marked in Fig.~\ref{fig:all the cases in the measurement} correspond to pairs of channels that vary in terms of their relative pathloss and spatial correlation, captured by $\alpha$ and $\rho$ defined in (\ref{eq:pathloss_param}) and (\ref{eq:spatial_corr_param}), respectively. The above table lists the empirical average of $\alpha$ and $\rho$ over all experiment realizations for each case.}
    \label{tab:channel_pathloss_correlation}
\end{table}

{
\begin{nrem}
\label{rem:channel_strengths}
For cases 1 through 3 in Fig.~\ref{fig:all the cases in the measurement}, we expect the relative pathloss difference between the two channels to be low by design, as the two RX antennas are fairly close together. In contrast, for cases 4 through 9, RX~2 is much farther from the TX than RX~1; hence, we expect $\alpha$ to be more negative by design. This is validated in Table~\ref{tab:channel_pathloss_correlation}, which lists the empirical average of $\alpha$ over all experiment realizations for each case.
\end{nrem}
}

\begin{nrem}[Channels with high spatial correlation]
\label{rem:high_corr}
In cases 1, 4 and 7, the channels are expected to be highly spatially correlated (i.e., small $\rho$) by design, as the two RXs are nearly in a straight line from the TX. {This is validated in Table~\ref{tab:channel_pathloss_correlation}, which lists the empirical average of $\rho$ over all experiment realizations. RSMA and NOMA are expected to provide large throughput gains over SDMA in these cases, as the common stream allows for the decoding and elimination of interference.} 
\end{nrem}

\begin{nrem}[Channels with low spatial correlation]
\label{rem:low_corr}
Similarly, in cases 3, 6 and 9, the spatial correlation between the two channels is expected to be relatively low (i.e., larger $\rho$) by design, {due to the larger angular separation between the RXs as seen by the TX. This is validated in Table~\ref{tab:channel_pathloss_correlation}. The gain of RSMA over SDMA is expected to be modest, with the common stream being allocated lower power relative to the cases in Remark~\ref{rem:high_corr}. NOMA is also unlikely to yield significant gains over SDMA; in fact, by forcibly imposing a common stream when the channels are not sufficiently aligned, the NOMA throughput performance can even be worse than SDMA.}
\end{nrem}

{
\subsection{RSMA v/s SDMA v/s NOMA: Sum Throughput Comparison}
\label{subsec:sum_tput_comparison}

\begin{figure}
    \centering
    \includegraphics[scale = 0.4]{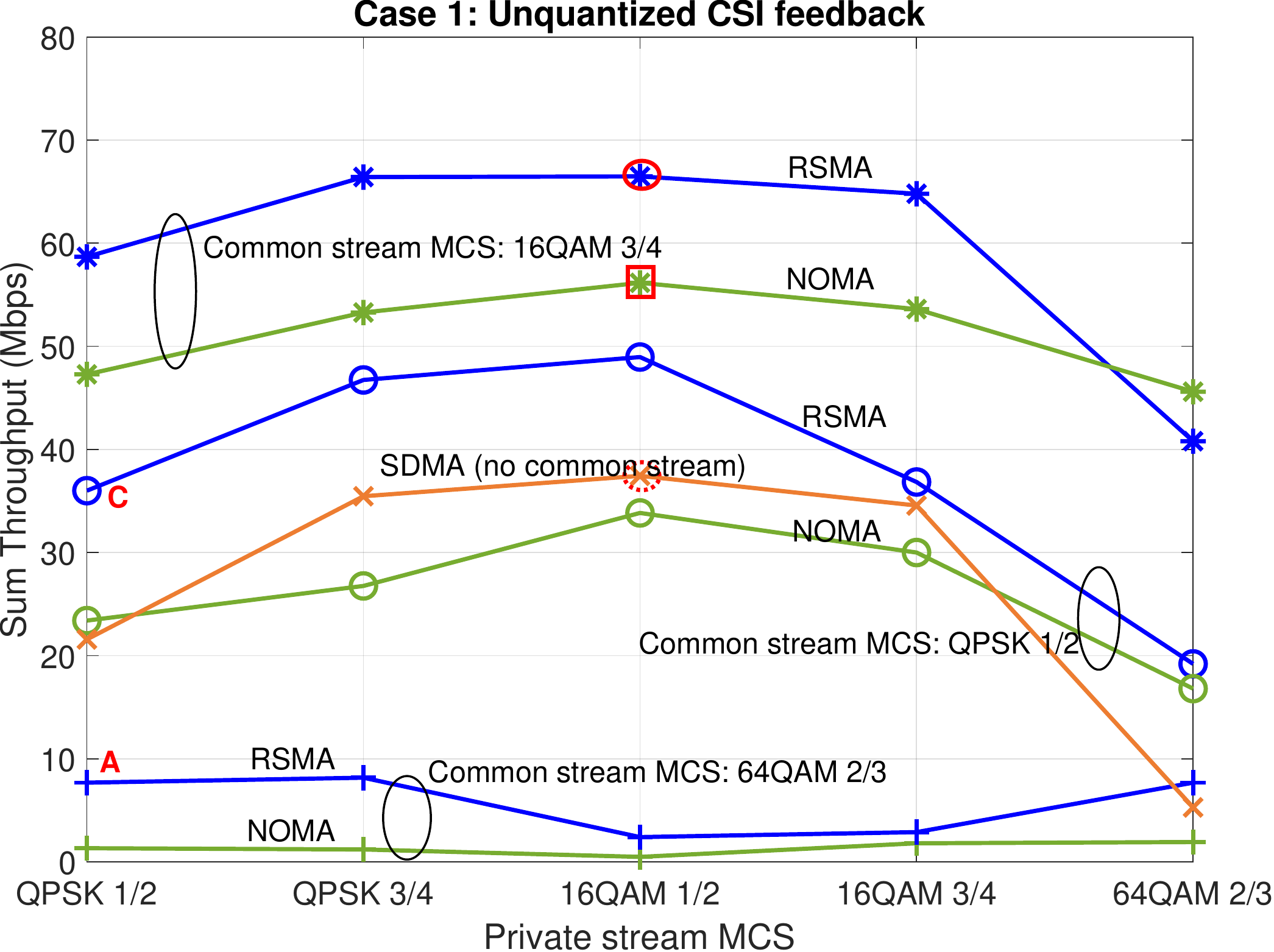}
    \caption{For Case~1, the large spatial correlation (i.e., small $\rho$, see Table~\ref{tab:channel_pathloss_correlation}) between the RX channels induces high mutual interference. Under these conditions, RSMA and NOMA can achieve significant throughput gains over SDMA with the help of the common stream, provided its MCS level is not too aggressive. The highest measured sum throughput above are: $66.48{\rm Mbps}$ for RSMA, $56.16{\rm Mbps}$ for NOMA, and $37.44{\rm Mbps}$ for SDMA.}
    \label{fig: tp curve of case 1}
\end{figure}

As an example measurement, the sum throughput performance for Case~1 as a function of the MCS levels is plotted in Fig.~\ref{fig: tp curve of case 1} with unquantized CSI feedback. For RSMA/SDMA, the MCS levels for the two private streams are chosen to be identical for ease of illustration\footnote{Since the two RXs are closely situated for this case, this is also an intuitively good choice. However, in general, the MCS levels of the private streams can be different depending on the channel realizations.}. For NOMA, the $x$-axis corresponds to the MCS of RX~1, as only RX~1 has a private stream. We make the following observations in Fig.~\ref{fig: tp curve of case 1}:
\begin{itemize}
    \item[1)] For a given common stream MCS level for RSMA/NOMA, the sum throughput increases with the private stream MCS level up to a point as expected, before dropping when the MCS level is too aggressive. The same behavior is observed for SDMA as well, which does not have a common stream.
    
    \item[2)] For a given private stream MCS level, a similar trend as above can also be seen w.r.t common stream MCS level for RSMA/NOMA. Specifically, the RSMA/NOMA sum throughput increases as the common stream MCS level is raised from QPSK rate $1/2$ (MCS index 2 in Table~\ref{tab: Mcs table}) to 16QAM rate $3/4$ (MCS index 5 in Table~\ref{tab: Mcs table}). However, the very next MCS level of 64QAM rate $3/4$ for the common stream is too aggressive and due to the resulting error propagation, the RSMA/NOMA sum throughput is lower than SDMA. To demonstrate this, Fig.~\ref{fig:superposed_constellation} shows the superposed received symbol constellation for RSMA at RX~1, corresponding to the MCS levels marked by points `C' and `A' in Fig.~\ref{fig: tp curve of case 1}. The MCS levels at point C being conservative, a QPSK constellation (private stream) superposed on another QPSK constellation (common stream) is clearly visible in Fig.~\ref{fig:superposed_constellation}a. In a similar vein, if the MCS levels at point A were \emph{not} aggressive, then we would expect to see 256 distinct symbol clusters in Fig.~\ref{fig:superposed_constellation}b, corresponding to a superposition of 64QAM (common stream) and QPSK (private stream) constellations. However, we only observe 81 distinct symbol clusters, which leads to high error rates for the common stream.
    \begin{figure}
    \centering
    \begin{subfigure}{0.49\textwidth}
        \centering
        \includegraphics[scale = 0.9]{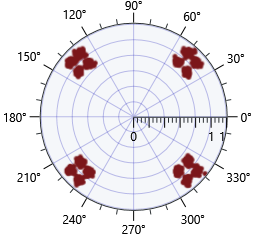}
        \caption{Common stream: QPSK $1/2$, private stream: QPSK $1/2$}
    \end{subfigure}
    \\
    \begin{subfigure}{0.49\textwidth}
        \centering
        \includegraphics[scale = 0.9]{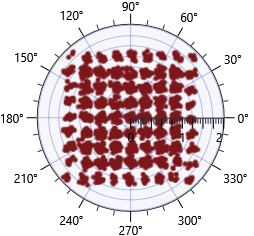}
        \caption{Common stream: 64-QAM $2/3$, private stream: QPSK $1/2$}
    \end{subfigure}
    \caption{Superposed received symbol constellation for RSMA at RX~1 for the MCS levels in Fig.~\ref{fig: tp curve of case 1} corresponding to: (a) point `C', which is an example of a conservative choice of MCS levels. Hence, the superposition of QPSK and QPSK constellations is clearly visible; (b) point `A', which is an aggressive choice of MCS for the common stream. Instead of $256$ distinct symbol clusters due to a superposition of 64-QAM and QPSK constellations, we only observe $81$ distinct symbol clusters, which leads to high error rates for the common stream.}
    \label{fig:superposed_constellation}
    \end{figure}

    \item[3)] Points 1 and 2 above highlight the role of MCS selection in realizing sum throughput gains for RSMA and NOMA over SDMA. The highest measured sum throughput in Fig.~\ref{fig: tp curve of case 1} is the empirical solution to $OP_{\rm mcs}^{\rm ma}$ in Table~\ref{tab:rsma_noma_sdma}, with the additional constraint that the MCS levels for the two private streams in RSMA/SDMA are identical. For RSMA, this value is equal to $66.48{\rm Mbps}$ (marked with a solid red circle); for NOMA, it is equal to $56.16{\rm Mbps}$ (red square), and for SDMA it is only $37.44{\rm Mbps}$ (dashed red circle).

    \item[4)] Finally, for all data points bar one, the RSMA sum throughput exceeds that of NOMA. Since RX 2's channel is not much weaker than RX 1 for Case 1 (see $\alpha$ in Table~\ref{tab:channel_pathloss_correlation}), RSMA provides additional throughput gain through the private stream of RX 2.
\end{itemize}

Fig.~\ref{fig: all tp in bar chart} plots the empirical solution for $OP_{\rm mcs}^{\rm ma}$, broken down in terms of the contributions of the common and private streams. We make the following observations:
\begin{itemize}
    \item[A.] \textbf{Unquantized CSI feedback}: 
    \begin{itemize}
      \item[1)] As expected for RSMA, the contribution of the common stream (green bars) diminishes with increasing $\rho$ (Remarks~\ref{rem:high_corr} and \ref{rem:low_corr}); in particular, for cases 1, 4 and 7, the common stream constitutes $51\%$, $62\%$ and $62\%$ of the sum throughput, respectively, whereas for cases 3, 6 and 9, this drops to $22\%$, $34\%$ and $17\%$, respectively. 

    \item[2)] In cases 1, 4 and 7, where the channels have high spatial correlation (low $\rho$, see Table~\ref{tab:channel_pathloss_correlation}), SDMA's sum throughput is significantly lower than those of RSMA and NOMA due to the high mutual interference. The gains of RSMA and NOMA in these cases stem from the decoding and elimination of interference through the common stream (Remark~\ref{rem:high_corr}). In particular, RSMA achieves gains of $82\%$, $103\%$ and $134\%$, respectively, over SDMA as shown by the up arrows in Fig.~\ref{fig: all tp in bar chart}). On the other hand, RSMA's gain over NOMA for these cases is quite modest ($23\%$ for Case 1, none for Case 4 and $19\%$ for Case 7).
    
    \item[3)] In cases 3, 6 and 9, where the channels have low spatial correlation (high $\rho$, see Table~\ref{tab:channel_pathloss_correlation}), RSMA's gain over SDMA is modest due to the relatively lower mutual interference. On the other hand, the NOMA sum throughput is significantly worse than both RSMA and SDMA. In particular, RSMA achieves gains of $38\%$, $45\%$, and $19\%$, respectively, for these three cases over NOMA.
    \end{itemize}

    \item[B.] \textbf{Quantized CSI feedback}:
    \begin{itemize}
        \item[1)] Unsurprisingly, all three multiple access schemes experience a sum throughput loss due to CSI quantization. However, SDMA and RSMA are more adversely affected than NOMA, especially in cases 4 through 9, where there is a significant pathloss difference between the two channels. For these cases, most of the throughput loss for RSMA and SDMA can be attributed to the private stream of RX 2 -- the weaker user. In contrast, the robustness of NOMA to CSI quantization stems from the fact that RX 2 does not have a private stream. Furthermore, due to the very low private stream throughput for RX~2, the RSMA throughput performance is similar to NOMA for these cases. 

        \item[2)] However, in the absence of a large pathloss difference between the two channels -- for instance, in cases 2 and 3 -- the private stream throughput for RX 2 is still quite high, despite the loss from CSI quantization. Under these conditions, NOMA with its missing private stream for RX~2 continues to perform worse than both RSMA and SDMA, mirroring the performance seen with unquantized CSI feedback.

        \item[3)] In all cases bar one, RSMA's gain over SDMA is much larger when compared to unquantized CSI feedback. This is because imperfect CSI has a more adverse effect on private stream throughputs due to the higher residual interference. Thus, due to the throughput contribution of the common stream, which stays pretty steady even with quantized CSI, RSMA's throughput loss is lower than that of SDMA.

    \end{itemize}
\end{itemize}

\begin{figure*}
\centering
    \begin{subfigure}{0.9\textwidth}
        \centering
        \includegraphics[width=\linewidth]{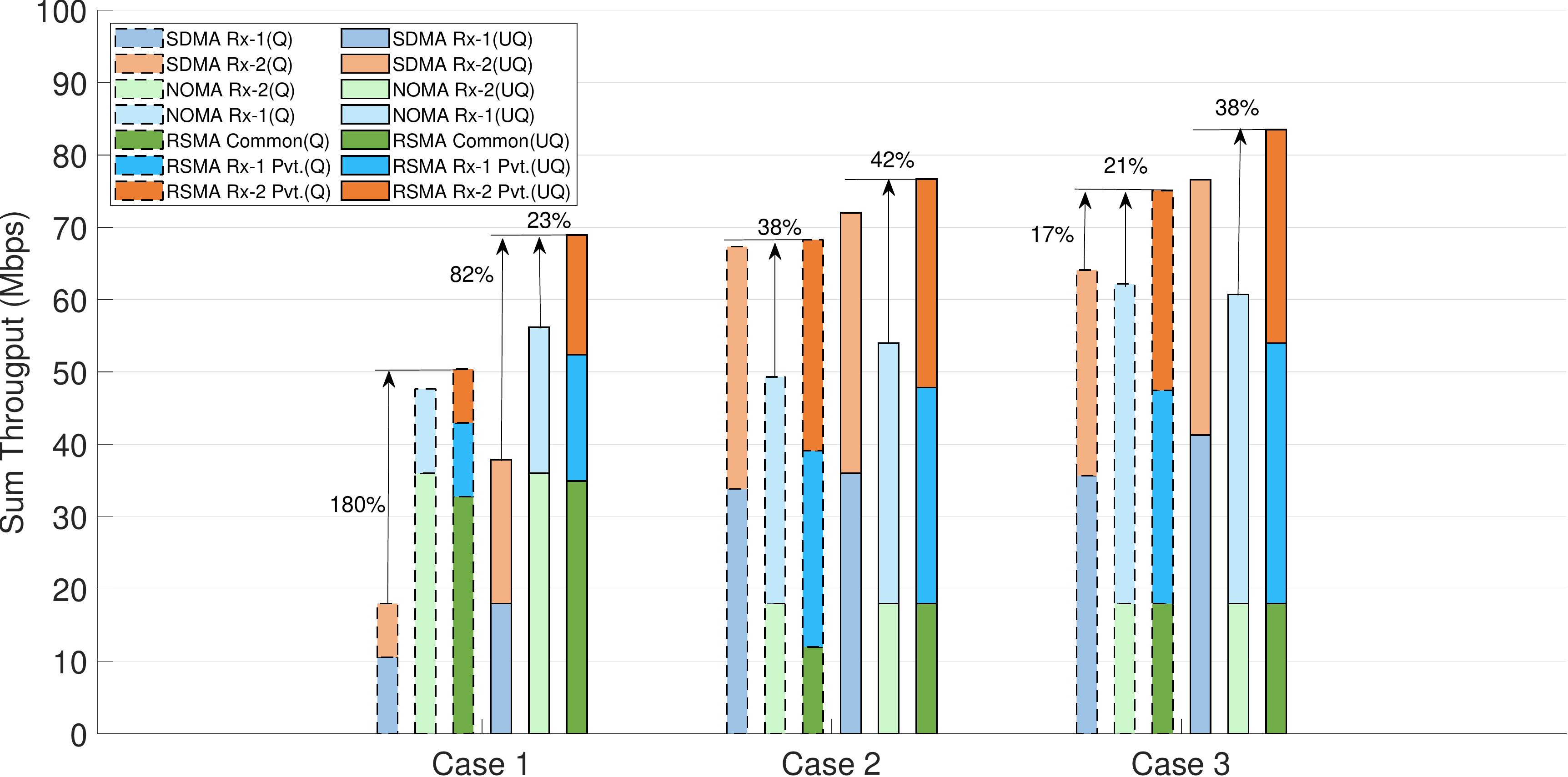}
        \caption{}
        \label{fig:sum tp 1-3}
    \end{subfigure}
    \begin{subfigure}{0.9\textwidth}
        \centering
        \includegraphics[width=\linewidth]{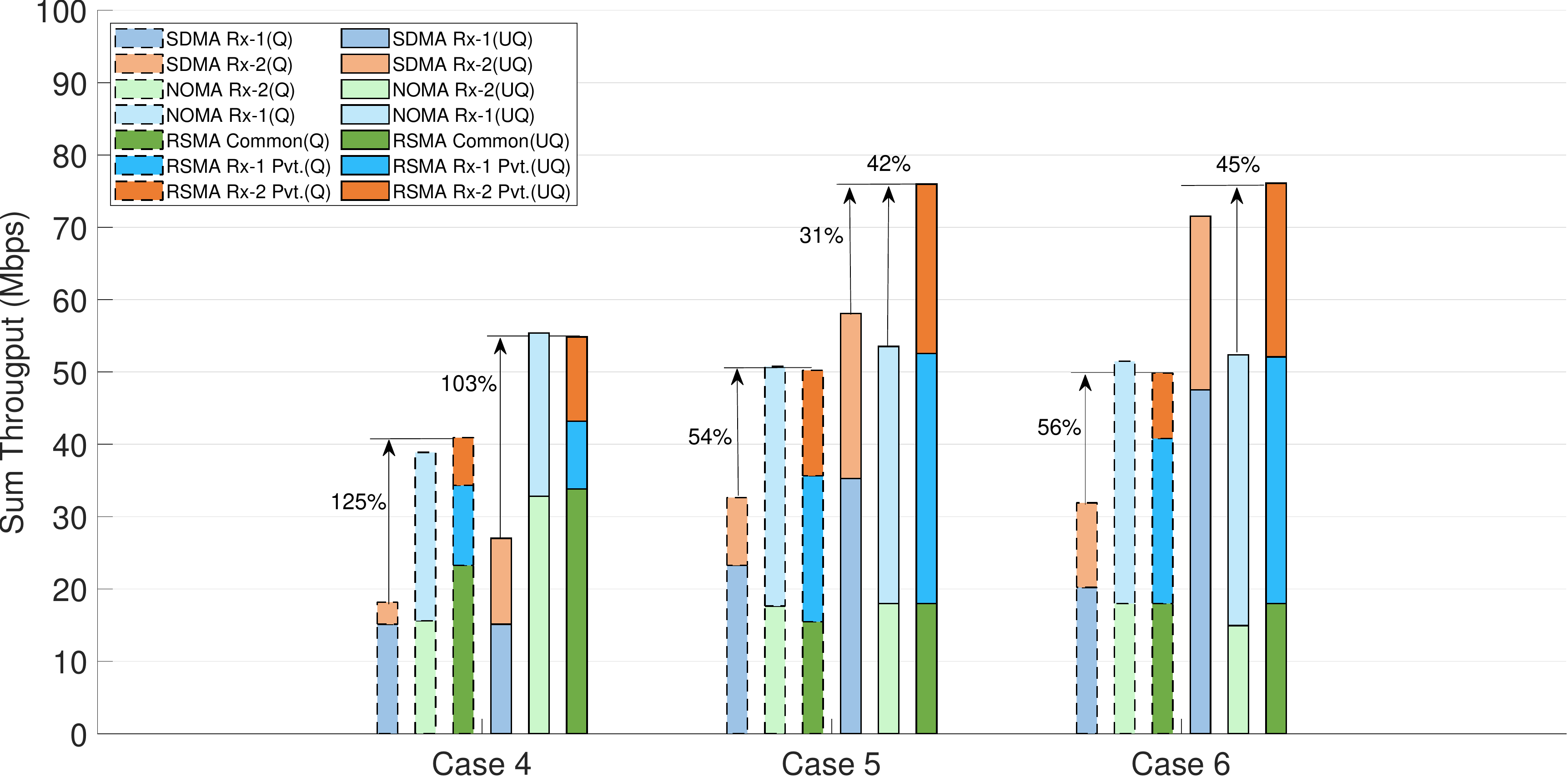}
        \caption{}
        \label{fig:sum tp 4-6}
    \end{subfigure}
    \caption{The maximum measured sum throughput under unquantized (UQ) and quantized (Q) CSI feedback for RSMA, SDMA and NOMA, which correspond to the empirical solutions of $OP_{\rm mcs}^{\rm RSMA}$, $OP_{\rm mcs}^{\rm SDMA}$ and $OP_{\rm mcs}^{\rm NOMA}$, respectively. The arrows indicate RSMA's gain w.r.t. SDMA and NOMA.} 
   \end{figure*}
   \begin{figure*}
\ContinuedFloat 
    \begin{subfigure}{0.9\textwidth}
        \centering
        \includegraphics[width=\linewidth]{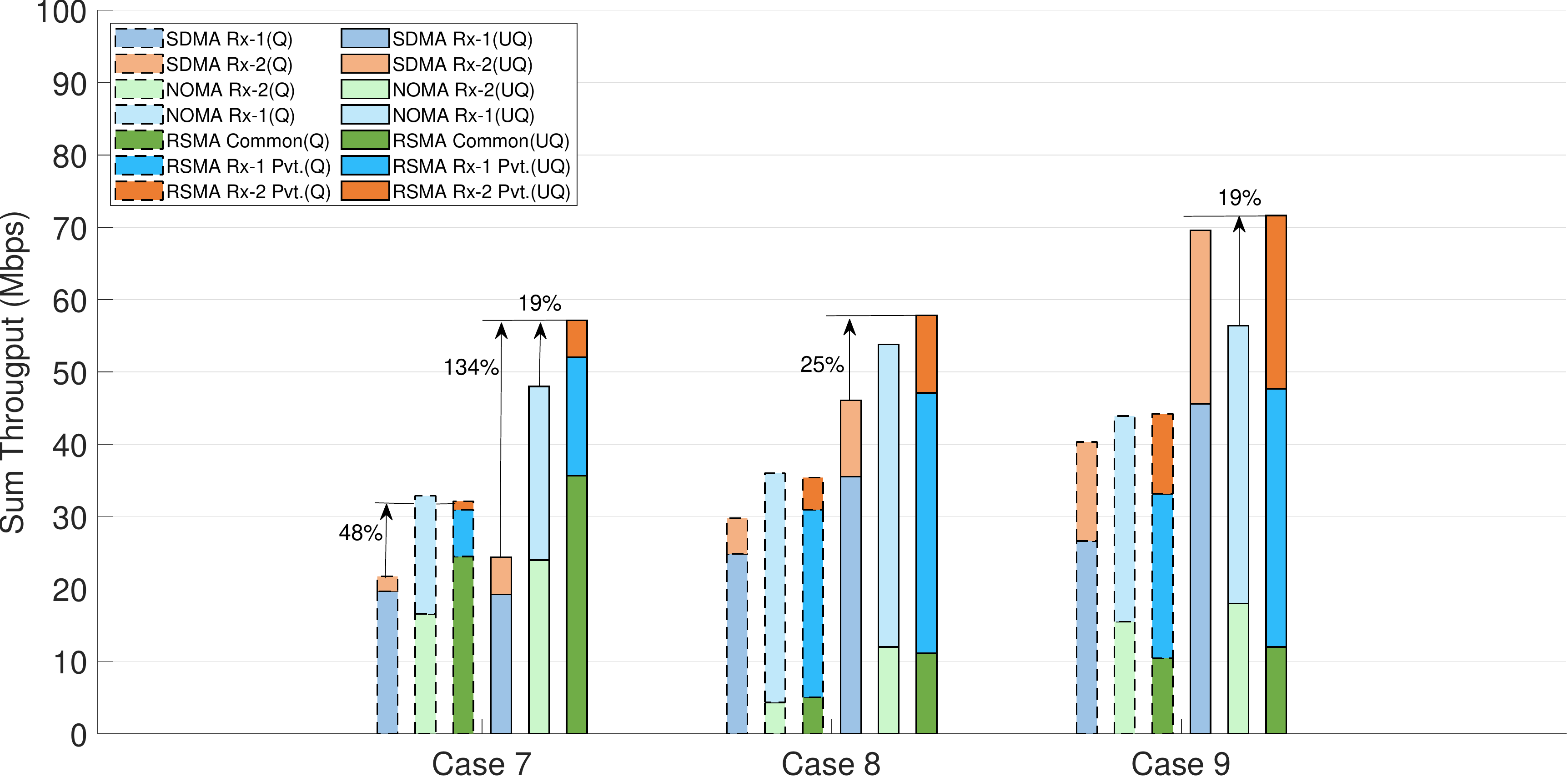}
        \caption{}
        \label{fig:sum tp 7-9}
    \end{subfigure}
   \caption{(contd.)}
    \label{fig: all tp in bar chart}
\end{figure*}
In summary, RSMA achieves the best throughput performance across (a) different RX channels that vary in their relative strength and spatial correlation, and (b) different CSI quality. In some cases, SDMA is effective while in others NOMA is effective. RSMA is effective in all the cases, and thus effectively bridges between both SDMA and NOMA.

\subsection{RSMA v/s SDMA v/s NOMA: Fairness Comparison}
\label{subsec:fairness_comparison}

In the previous subsection, we saw that for RSMA, the private stream throughput of RX~2 is severely diminished when there is a large pathloss difference between the two channels and imperfect CSI (bullet point B1). To increase RX 2's total throughput in such cases, a larger fraction ($> 0.5$) of the common stream can be allocated to it. This would not alter the RSMA sum throughput\footnote{For precisely this reason, the allocation of the common stream between the two RXs is not mentioned when discussing the sum throughput results in Section~\ref{subsec:sum_tput_comparison}.}, but would achieve better (max-min) fairness \cite{mao2022fundmental}. 

In this subsection, we experimentally verify this possibility of improved fairness afforded by RSMA. This is done in the following manner for each case in Fig.~\ref{fig: all tp in bar chart}:
\begin{itemize}
    \item[S1)] if $R_1(\nbP_{\rm wmmse}^{\rm RSMA}) - R_2(\nbP_{\rm wmmse}^{\rm RSMA}) \leq R_c (\nbP_{\rm wmmse}^{\rm RSMA})$, then the common stream is partitioned such that $R_1 (\nbP_{\rm wmmse}^{\rm RSMA}) + (W_{c,1}/W_c) R_c (\nbP_{\rm wmmse}^{\rm RSMA}) = R_2 (\nbP_{\rm wmmse}^{\rm RSMA}) + (W_{c,2}/W_c) R_c (\nbP_{\rm wmmse}^{\rm RSMA})$ to ensure that both RXs have the same throughput;
    \item[S2)] otherwise, when $R_1 (\nbP_{\rm wmmse}^{\rm RSMA}) - R_2 (\nbP_{\rm wmmse}^{\rm RSMA}) > R_c (\nbP_{\rm wmmse}^{\rm RSMA})$, the entire common stream is allocated to RX~2. This is similar to NOMA, albeit RX~2 has a private stream as well.
\end{itemize}
\begin{nrem}[Fairness while maximizing throughput]
    User fairness in RSMA/SDMA/NOMA is often framed as an optimization problem, where the objective is to maximize the minimum user throughput through precoder design. Hence, the optimal precoders for fairness are, in general, different from the optimal precoders for sum throughput maximization. The fairness strategy in S1 and S2 differs from this viewpoint by trying to achieve fairness without changing the precoders. In other words, we are trying to achieve fairness while still maximizing the sum throughput. This ability to realize both objectives simultaneously is a unique feature of RSMA. S1 and S2 cannot be realized for NOMA and SDMA because the former does not have a private stream for RX 2 and the latter does not have a common stream.
\end{nrem}

Fig.~\ref{fig:tp vs min user} compares the relationship between the sum throughput and the minimum throughput for RSMA (subject to S1 and S2), SDMA and NOMA. The case number is indicated beside each data point in the same color. The dashed $y=2x$ line represents max-min fairness and the region above it is feasible for all three multiple access schemes. Thus, points that are closer (in terms of Euclidean distance) to the dashed line represent \emph{fairer} outcomes. Based on this insight, we make the following remarks:
\begin{itemize}
    \item When S1 is true, the corresponding RSMA point lies on the dashed line (i.e., achieves max-min fairness), as seen for all cases bar one in Figs.~\ref{fig:tp vs min user}a and b. Otherwise, for S2, the RSMA point lies \emph{as close as possible} above the dashed line, as seen for case 8 in Figs.~\ref{fig:tp vs min user}a and b.

    \item In Fig.~\ref{fig:tp vs min user}a, SDMA also achieves approximate fairness for some cases (namely 1, 2, 3 and 4), as does NOMA (case 7). However, the RSMA points achieve fairness at a higher sum for these cases lie to the \emph{northeast} of the corresponding SDMA/NOMA points, which indicates fairness at a higher sum throughput.
    
\end{itemize}
In summary, RSMA is capable of achieving better fairness at a higher sum throughput than both SDMA and NOMA, regardless of CSI quality, through a judicious allocation of the common stream to the weaker RX.


\begin{figure}[h!]
\begin{subfigure}{0.49\textwidth}
 \centering
    \includegraphics[width=\linewidth]{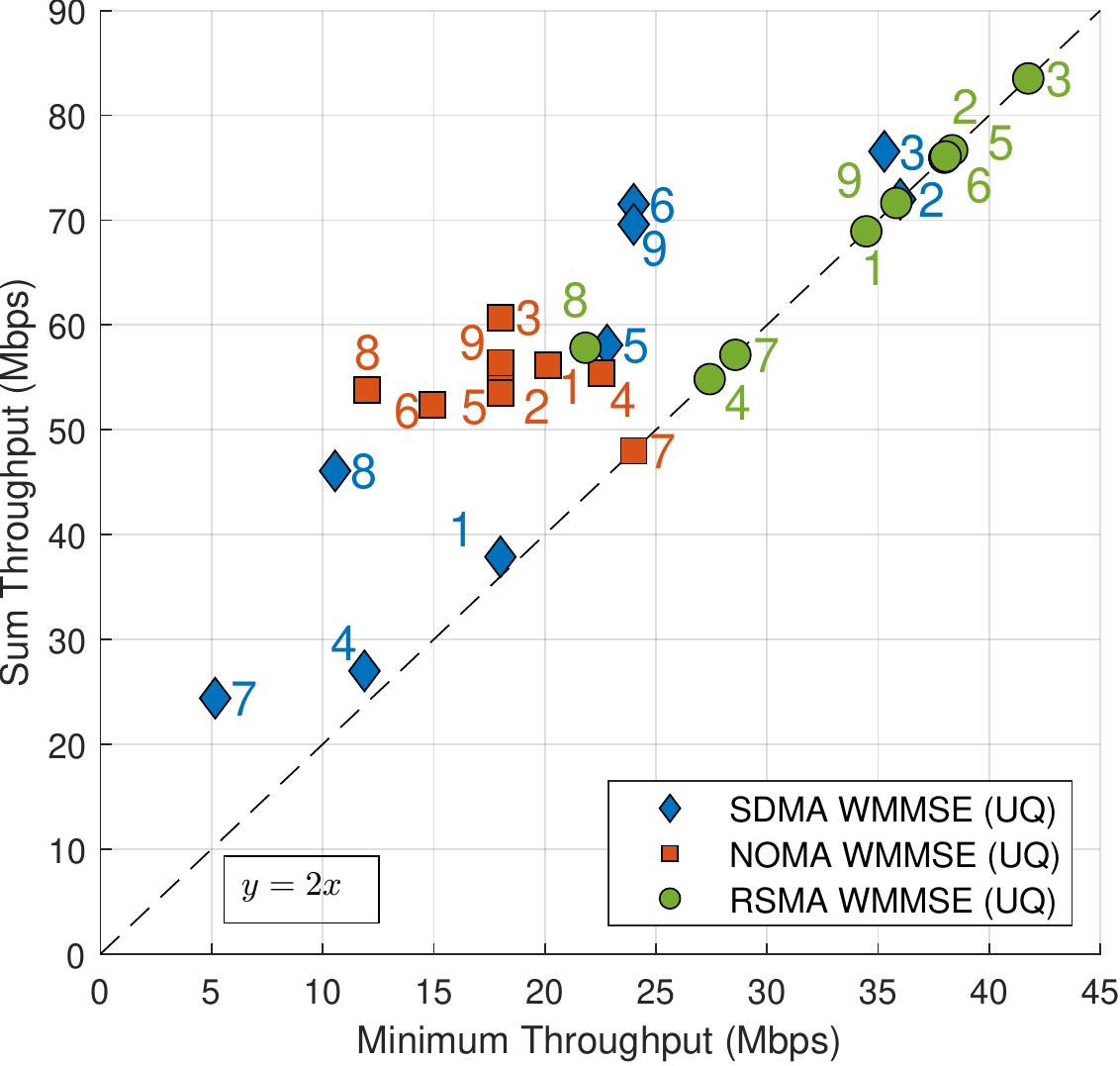}
    \caption{Unquantized CSI feedback}
    \label{fig:tp vs min wmmse unquantized}
    \end{subfigure}
    \begin{subfigure}{.49\textwidth}
    \centering
    \includegraphics[width=\linewidth]{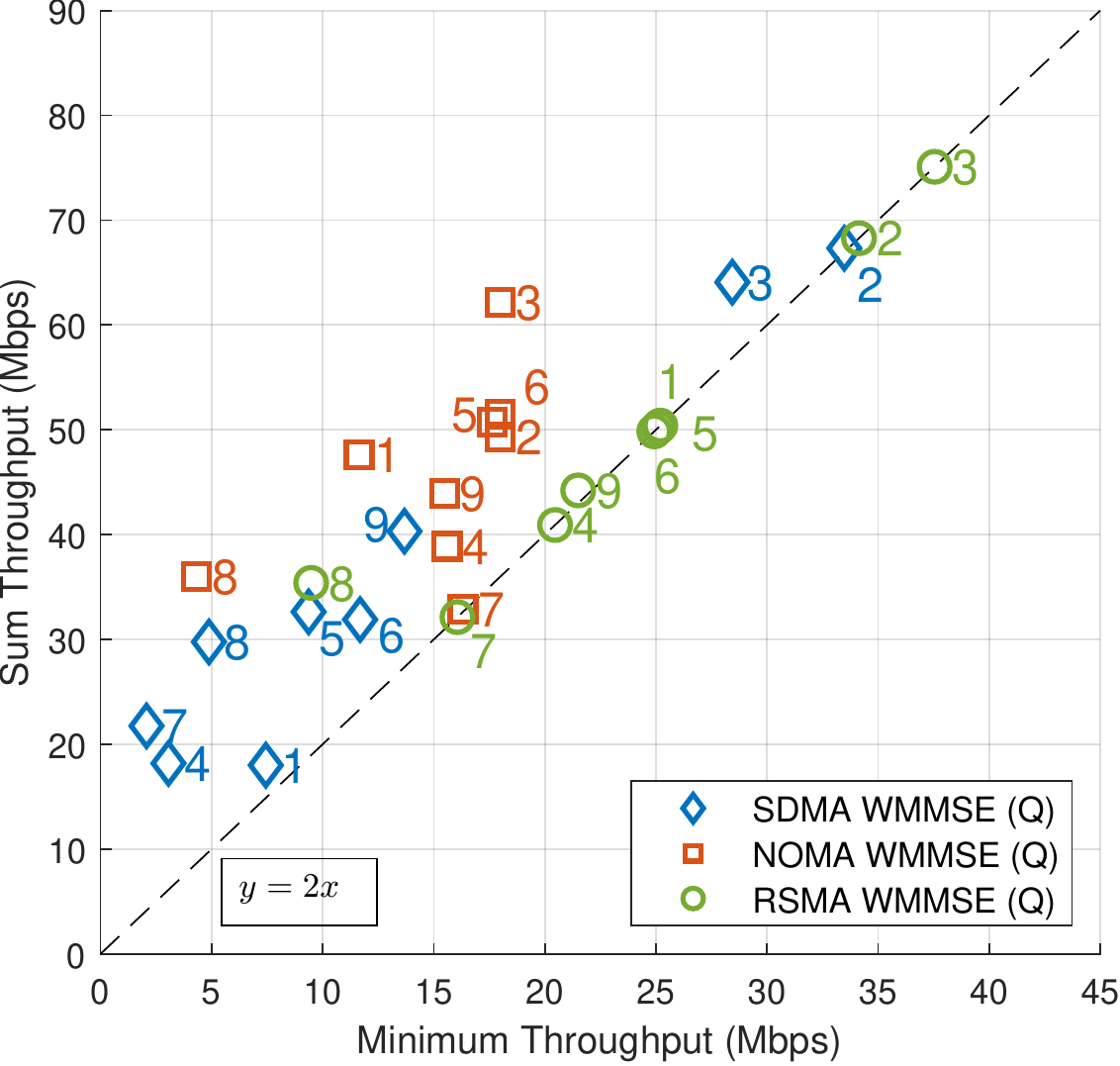}
    \caption{Quantized CSI feedback}
    \label{fig:tp vs min wmmse quantized}
    \end{subfigure}
    \caption{Fairness comparison between RSMA (subject to S1 and S2), SDMA and NOMA. The number beside each data point indicates the measurement case from Fig.~\ref{fig:all the cases in the measurement}. The dashed line ($y=2x$) corresponds to max-min fairness and points that are closer (in terms of Euclidean distance) to this line represent fairer outcomes.}
    \label{fig:tp vs min user}
\end{figure}

}

\section{Summary}
\label{sec:summary}
In this paper, we realized the first-ever RSMA prototype using SDRs in order to experimentally validate its performance gains promised in the theoretical literature. To this effect, we measured the throughput and fairness performance of RSMA, SDMA and NOMA using our prototype for the two-user MISO scenario over (a) different pairs of channels varying in terms of their relative strength and spatial correlation and (b) different CSI quality. Across these different cases, we observed that RSMA achieved the highest sum throughput, whereas SDMA and NOMA were each effective only in some cases. We also observed that RSMA achieved better fairness at a higher sum throughput than both SDMA and NOMA. As the first-ever experimental comparison between RSMA, SDMA and NOMA,
our efforts in this paper address the pressing question of which multiple access technique is best equipped to deliver the performance enhancements expected from next-generation networks like 6G. Thus, the superior performance of RSMA in these experiments provides the impetus for further experimental efforts at a larger scale and in more challenging operating environments.

\section*{Acknowledgments}
We thank Onur Dizdar for his valuable inputs on Polar code implementation.

\appendices
\section{CSI Quantization Technique}
\label{appendix:A}



The quantized wideband CSI, $\hat{\mathbf{h}}^{\rm Q}_{i} := [\hat{h}_{i1}^{\rm Q} ~ \hat{h}_{i2}^{\rm Q}]$, is obtained from $\hat{\mathbf{h}}_{i} = [\hat{h}_{i1} ~ \hat{h}_{i2}]$, as follows:
\begin{itemize}
    \item [(a)] The \emph{linear scaler} is evaluated as follows:
    \begin{align}
        \label{eq: CSI scaler linear}
     M_h^{\mathrm{lin}} &:= \frac{m_h}{10^{M_h/20}} \\
        \label{eq:CSI encoding find max}
     \mbox{where}~  m_{h} &:= \max_{i,l} \{|\mathrm{Re}(\hat{h}_{il})|, |\mathrm{Im}(\hat{h}_{il})|\} ~ (i,l \in \{1,2\}), \\
     \label{eq:CSI quantized scale}
     \mbox{and}~M_h &:= \min\biggl\{7,\Bigl\lfloor 20\log_{10}(m_h)\Bigr\rfloor\biggr\}.
     \end{align}
    $M_h$ in (\ref{eq:CSI quantized scale}) is known as the scaling ratio and is quantized to three bits for feedback.

    \item [(b)] Using (\ref{eq: CSI scaler linear}), the real and imaginary parts $\hat{h}_{ij}$ are quantized using $N_b=4$ bits in our prototype as follows.
    \begin{align}
    \label{eq:CSI quantization bits real}
    {\rm Re}(\hat{h}_{ij}^{\rm Q}) &:= \mathrm{round}\biggl(\frac{\mathrm{Re}(\hat{h}_{ij})}{M_h^{\mathrm{lin}} }(2^{N_b-1}-1)\biggr) \\
    \label{eq:CSI quantization bits imag}    
    {\rm Im}(\hat{h}_{ij}^{\rm Q}) &:= \mathrm{round}\biggl(\frac{\mathrm{Im}(\hat{h}_{ij} )}{M_h^{\mathrm{lin}} }(2^{N_b-1}-1)\biggr)
    \end{align}
\end{itemize}
The feedback overhead for the entire CSI equals $3 + 2N_bil$ bits, where the first term corresponds to the scaling ratio, and the factor of two in the second term accounts for the real and imaginary parts of $\hat{h}_{ij}^{\rm Q}$. For our prototype, this value equals $35$ bits, as opposed to the $128il = 512$ bits without quantization.

At the TX, the CSI estimates, denoted by $\hat{h}_{il}^{\rm TX}$, are recovered as follows:
\begin{align}
    \label{eq:CSI recovery real}
        \hat{h}_{il}^{\rm TX} := \frac{\hat{h}_{il}^{\rm Q}}{10^{M_h/20}}~ (i,l \in \{1,2\}).
\end{align}

\bibliographystyle{IEEEtran}

\end{document}